\documentclass[prc,twocolumn,secnumroman,tightenlines,amssymb,aps,nobibnotes,superscriptaddress,balancelastpage,floatfix,nofootinbib,linenumbers]{revtex4}

\usepackage{amsmath}
\usepackage{lipsum}
\usepackage{graphicx}
\usepackage{multirow}           
\usepackage{array}

\DeclareRobustCommand\substyle{\name@idx{document substyle}}
\DeclareRobustCommand\classoption{\name@idx{document class option}}
\DeclareRobustCommand\classname{\name@idx{document class}}
\def\name@idx#1#2{{\ttfamily#2}
\index{#2\space#1=\string\ttt{#2}\space#1}\index{#1>#2=\string\ttt{#2}}}

\begin{document}








\title{Particle emission from $^{40}$Ar nuclei excited in fixed-target $^{208}$Pb+$^{40}$Ar UPC at the LHC}

\author{P.~Jucha}
\email{Pawel.Jucha@ifj.edu.pl}
\affiliation{Institute of Nuclear Physics PAN, ul.\,Radzikowskiego 152, Pl-31342 Krak\'ow, Poland}

\author{M.~K\l{}usek-Gawenda}
\email{Mariola.Klusek@ifj.edu.pl}
\affiliation{Institute of Nuclear Physics PAN, ul.\,Radzikowskiego 152, Pl-31342 Krak\'ow, Poland}

\author{K.~Mazurek}
\email{Katarzyna.Mazurek@ifj.edu.pl}
\affiliation{Institute of Nuclear Physics PAN, ul.\,Radzikowskiego 152, Pl-31342 Krak\'ow, Poland}

\author{L. L. Pappalardo}
\email{Luciano.Libero.Pappalardo@unife.it }
\affiliation{Universit\'a degli Studi di Ferrara, Dipartimento di Fisica e Scienze della Terra, Via Saragat 1, 44122 Ferrara, Italy}
\affiliation{INFN Sezione di Ferrara, Via Saragat 1, 44122 Ferrara, Italy}

\author{K.~Pysz}
\email{Krzysztof.Pysz@ifj.edu.pl}
\affiliation{Institute of Nuclear Physics PAN, ul.\,Radzikowskiego 152, Pl-31342 Krak\'ow, Poland}

\author{A.~Szczurek}
\email{Antoni.Szczurek@ifj.edu.pl}
\affiliation{Institute of Nuclear Physics PAN, ul.\,Radzikowskiego 152, Pl-31342 Krak\'ow, Poland}
\affiliation{Institute of Physics, Faculty of Exact and Technical Sciences, University of Rzesz\'ow, Pigonia 1 St., 35-310 Rzesz\'ow, Poland}

\date{\today}

\begin{abstract}
Ultraperipheral collisions of heavy ions, dominated by the electromagnetic interaction, allow to investigate
the photoproduction of baryons and mesons from the excited nuclei. At LHCb, thanks to the SMOG2 system,
these studies can also be conducted in fixed-target collisions, exploiting a variety of target gases ranging from $^{4}$He to Xe. In the present work, we adapted our photonuclear model and the Giesen-Boltzmann-Uehling-Uhlenbeck (GiBUU) generator to study photoproduction of protons, neutrons as well as
pions and kaons from excited $^{40}$Ar nuclei in fixed-target ultraperipheral collisions with the LHC
$^{208}$Pb beam. We present predictions for the total and differential photoproduction cross sections in $p$
and $p_T$, both in the full phase-space and within the geometrical acceptance of the LHCb detector.
\end{abstract}


%


\maketitle


\section{Introduction}
\label{Sect.I}
Heavy-ion collisions at ultrarelativistic energies give the opportunity to
investigate the nucleon-nucleon interaction. In the case of ultraperipheral collisions (UPC), only the
electromagnetic fields generated by the colliding nuclei are responsible for their mutual excitation and the subsequent emission of hadrons. In our previous articles \cite{Klusek-Gawenda:2013ema, Jucha2025, Jucha2026,Jucha2025a}, the phenomenon of producing protons and neutrons in the $\gamma+A$ reaction in UPC was discussed. No meson production was studied so far in photoproduction at $E_\gamma$~$>$~1~GeV. At energies below 1~GeV, pion production was discussed e.g., in \cite{KRUSCHE_2005}. The kaon photoproduction was studied in $\gamma p$ processes by the SAPHIRE Collaboration \cite{SAPIHIRE_1998}.

The photoproduction of hadrons is an interesting phenomenon, although not yet well explored experimentally, especially at high photon energies.
In particular, due to experimental constraints, the study of the dependence on the photon energy is limited to about 1 GeV. Higher photon energies are available in principle in UPC's of heavy ions. In UPC at the Large Hadron Collider (LHC) energies, particles from the mutually excited nuclei are emitted
in the very forward/backward directions and can only be revealed by detectors placed at very-high (positive
or negative) pseudorapidity. So far, only the photoproduction of neutrons and protons has been studied at the LHC by
exploiting the Zero Degree Calorimeter (ZDC) of the ALICE experiment \cite{ALICE502a,ALICE_protons}. The study of pions and kaons
photoproduction was not possible with this experimental configuration. Here we wish to explore a new possibility by studying both baryon and meson
photoproduction in fixed-target UPC’s exploiting the SMOG2 target system of LHCb. In this case, the
ultrarelativistic $^{208}$Pb nuclei from the LHC beam can be treated as intense sources of associated
photons that excite the nuclei of gaseous targets, which can span from $^{4}$He to Xe. The emitted particles are then detected and identified by the LHCb detector.
In the present exploratory studies, we make use of the Giessen Boltzmann-Uehling-Uhlenbeck (GiBUU) generator,
which allows to simulate photon-nucleus collisions
with several hadronic output channels, in combination with the GEMINI++ de-excitation
code.

The procedure used in our calculation is similar to the method described in detail in \cite{Klusek-Gawenda:2013ema, Jucha2025, Jucha2026}. The hybrid approach consists of three stages:
(1) generation of the photon flux within the Equivalent Photon Approximation \cite{b-space-EPA}, (2) simulation of the $\gamma$-nucleus collision with GiBUU \cite{GiBUU} and (3) de-excitation of the hot remnants with GEMINI++\cite{gemini}. This method turned out to be very powerful in the predictions of the cross
section for multiple emissions of neutrons and protons
\cite{Klusek-Gawenda:2013ema, Jucha2025, Jucha2026}.

The present studies are focused on $^{208}$Pb+$^{40}$Ar ultraperipheral fixed-target collisions at
LHCb.\footnote{For comparison, we also present results for
hypothetical $^{208}$Pb+$^{208}$Pb fixed-target collisions, which presently cannot be performed at the
LHC.}
In addition to the
emission of nucleons (protons and neutrons), already studied in the past, the emission of light mesons ($\pi^{\pm,0}$, K$^{\pm}$, K$^0$ and $\bar{\text{K}}^0$) is here investigated for the first time.

\section{Electromagnetic Dissociation}
\label{Sect.II}
In ultrarelativistic collisions, the colliding nuclei are the source of the electromagnetic fields. Our previous papers \cite{Jucha2025, Jucha2026}
presented predictions
for photoproduction cross sections in the ultraperipheral $^{208}$Pb+$^{208}$Pb reaction, where both nuclei move at a velocity close to the speed of light. In contrast, the present manuscript considers the case of a $^{208}$Pb beam colliding with a $^{40}$Ar target at rest. Thus, the main idea is similar to that in our previous papers and the Equivalent Photon Approximation is used to predict the energy distribution of the photons emitted by the incoming beam particle.

The photon flux is given by
\begin{equation}
\begin{split}
    N(E_\gamma, b) &= \frac{Z^2 \alpha_\text{em}}{\pi^2 \beta^2}\frac{1}{E_\gamma b^2} \\ &\times \left|
            \int d\chi \hspace{3pt} \chi^2 
            \frac{F(\frac{\chi^2+u^2}{b^2})}{\chi^2+u^2}J_1(\chi) \right|^2 ,
    \end{split}\label{nomega}
\end{equation}

\noindent where $Z^2$ is the square of the beam atomic number, $\alpha_{\text{em}}$ stands for the fine-structure constant, $b$ represents the impact parameter, and $E_\gamma$ is the photon energy. The formula includes the dependence on the nuclear form factor $F(\frac{\chi^2+u^2}{b^2})$, which in our calculation is obtained as the Fourier transform of the nucleus charge density using the Fermi distribution, with $J_1$ denoting Bessel function of the first kind. The auxiliary variables $\chi = k_\perp b$, where $k_\perp$ is the photon transverse momentum, and $u = \frac{E_\gamma b}{\gamma\beta}$ are used to simplify numerical calculations.

\begin{figure}[!bt]
	\centering
	\includegraphics[width=0.45\textwidth]{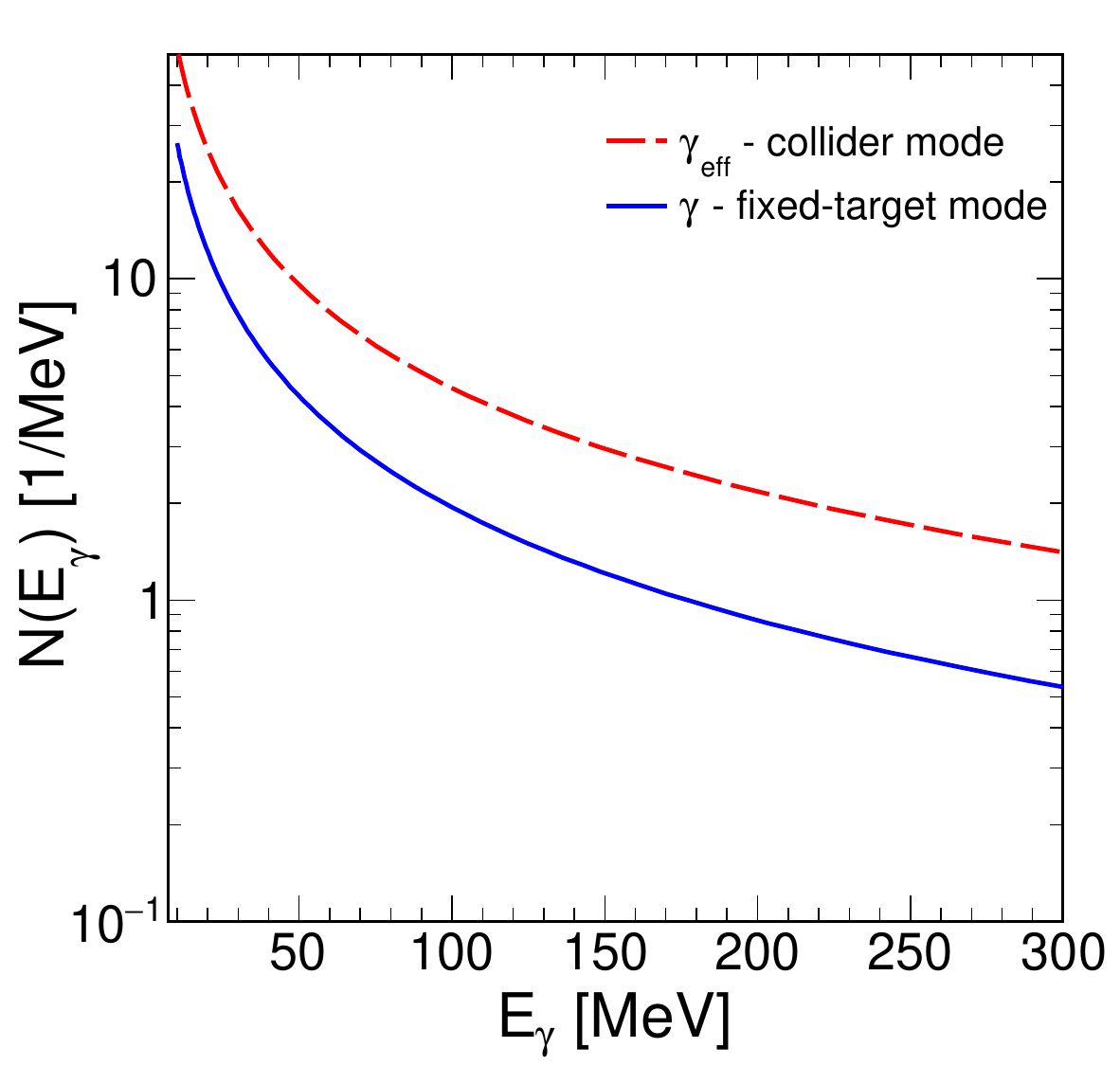}
	\caption{Photon flux (Eq.~\ref{nomega}) as a function of the photon energy $E_\gamma$ for lead nuclei with beam energy per nucleon $E_{beam}/A$~=~2.51~TeV integrated over the impact parameter $b$. The red dashed line represents the flux obtained with the effective Lorentz factor $\gamma_{eff}$, which is valid for the collider mode. The blue line shows the flux distribution in the rest frame of the target nucleus for the case of fixed-target collisions. In this case,
    the plain Lorentz factor $\gamma$ is used.}
	\label{fig:flux_comparison}
\end{figure}

The main difference between the calculation of the photon flux coming from two moving sources and the fixed-target collisions case is shown in Fig.~\ref{fig:flux_comparison}. The effective Lorentz factor $\gamma_\text{eff}$ (Eq.~\ref{gammaef}), which is to be used for the case of beam-beam collisions, depends on the velocity of both participants and,
by definition, is larger than the standard Lorentz factor $\gamma$.

\begin{equation}
 \gamma_\text{eff} = 2\gamma^2-1,\quad   \gamma = \frac{1}{\sqrt{1-\beta^2}}. \label{gammaef}
\end{equation}

As a result, the photon flux for the
$^{208}$Pb+$^{40}$Ar fixed-target case, based on the standard Lorentz factor $\gamma$, is significantly smaller than
that for the $^{208}$Pb+$^{208}$Pb reaction from colliding beams, based on the effective Lorentz factor
$\gamma_\text{eff}$.

Using the photon flux given in Eq.~\ref{nomega}, we can write the equation for the total cross section in UPC:
\begin{equation}
\label{UPC_cs}
    \sigma_\text{tot} = \int dE_\gamma \cdot db \cdot 2\pi b \cdot N(E_\gamma, b) \sigma_{\gamma A\to X}(E_\gamma) \cdot \text{exp}[-m(b)].
\end{equation}

\noindent 
Here,  $\sigma_{\gamma A\to X}(E_\gamma)$ denotes the cross section for the $\gamma A\to X $ process of interest as a function of the photon energy and $m(b)$ represents the mean number of absorbed photons, given by
\begin{equation}
\label{mean_photon}
	m(b) = \int dE_\gamma N(E_\gamma, b) \sigma_\text{abs}(E_\gamma) .
\end{equation}

\begin{figure}[!bt]
\hskip -1.22cm
\includegraphics[width=0.55\textwidth]{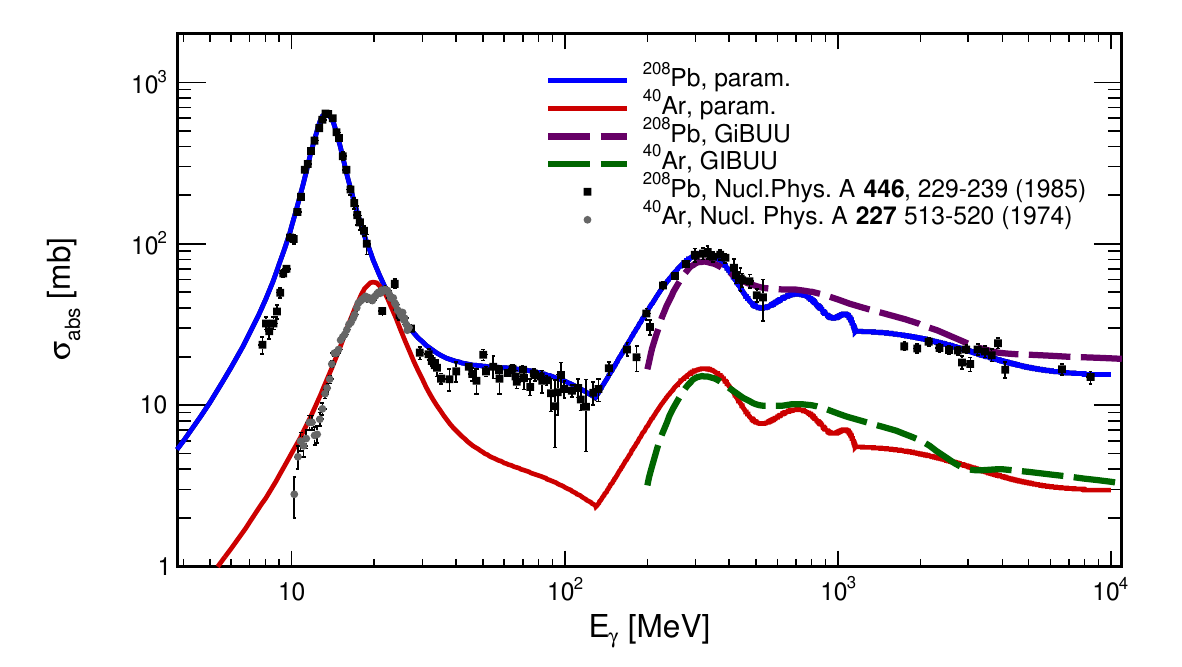}
\caption{Photoabsorption cross section as a
function of $E_\gamma$. The blue and red lines are related to the parametrization of the cross section for $^{208}$Pb and $^{40}$Ar targets, respectively. The green and purple line depicts results obtained from GiBUU. The black squares are experimental results from \cite{AHRENS_1985} and the gray dots from \cite{VEYSSIERE_1974}.}
\label{fig:total_cs}
\end{figure}

\noindent The photoabsorption cross section $\sigma_\text{abs}$ 
for $^{208}$Pb is known experimentally in a wide photon-energy range. For the $\gamma$+$^{40}$Ar reaction, only the Giant Dipole Resonance region has been measured up to $E_{\gamma}$=25~MeV (Fig.~\ref{fig:total_cs}). In our calculations, the $^{208}$Pb photoabsorption cross section has been parametrized and rescaled applying the appropriate $Z$- and $A$-dependencies to obtain results for the $^{40}$Ar target. In the Giant Resonance Region, specific parameters for $^{40}$Ar nucleus were adopted from the available experimental literature~\cite{VEYSSIERE_1974}, ensuring consistency with the measured low-energy photoabsorption cross section.

Our parametrization is in good agreement with the available experimental results \cite{VEYSSIERE_1974, AHRENS_1985}. In our previous work,
the GiBUU approach has been used to simulate
neutrons and protons production in $\gamma$-nuclei collisions at photon energies $E_{\gamma}$ higher than
200 MeV and 580 MeV, respectively. Our parametrization and the GiBUU predictions of the photoabsorption cross section for $^{208}$Pb and $^{40}$Ar cases agree quite well.

The total cross section of electromagnetic dissociation (EMD) of target nuclei can be calculated from the Poisson formula, taking into account all possible cases where the nucleus absorbs one or multiple photons. The corresponding formula can be written as
\begin{equation}
	\sigma_{EMD} = \int db \cdot 2\pi b \cdot \big(1 - \text{exp}[-m(b)] \big) \label{sigmaemd}.
\end{equation} 

\noindent 
The EMD cross section for a hypothetical Pb+Pb fixed-target reaction,
considered here only for reference, is 84.7~b. This value is significantly larger than that of the corresponding
cross section for the Pb+Ar fixed-target 
reaction here under study, which is equal to 13.3 b. This difference arises from the
different numbers of nucleons in the two target nuclei (208 against 40). However, the dependence of the cross section on the atomic mass number is found to slightly deviate from bare proportionality.

 A microscopic simulation of nucleus-nucleus collision with the electromagnetic dissociation is presently not available. For this reason, as it was already explained in \cite{Jucha2026}, the commonly recognized GiBUU model \cite{GiBUU} is utilized for current studies.
This model, originally developed to describe heavy-ion and nucleon-nucleon collisions, was extended to include real and virtual photon-nucleus interactions. The description of the strategy and the initial parameters applied for GiBUU simulations
performed for the current work are provided in~\cite{Jucha2026}.

\section{Nucleon Emission}
\label{Sect.III}
In UPC both low- and high-energy $\gamma$-nucleus collisions occur. The high-energy collisions are treated with the GiBUU generator \cite{GiBUU}. Here, the electromagnetic dissociation of the colliding nuclei is described by a two-step process. The primary emission of hadrons provided by GiBUU leads to remnants whose excitation energy is subsequently treated with GEMINI++. The estimation of the excitation energy is discussed in detail in \cite{Jucha2025}. Here we present only a sketchy description of the method.

\begin{figure}[!bt]
\setlength{\unitlength}{0.1\textwidth}
\begin{picture}(4,5.9)
\put(-0.5,2.8){\includegraphics[width=0.5\textwidth]{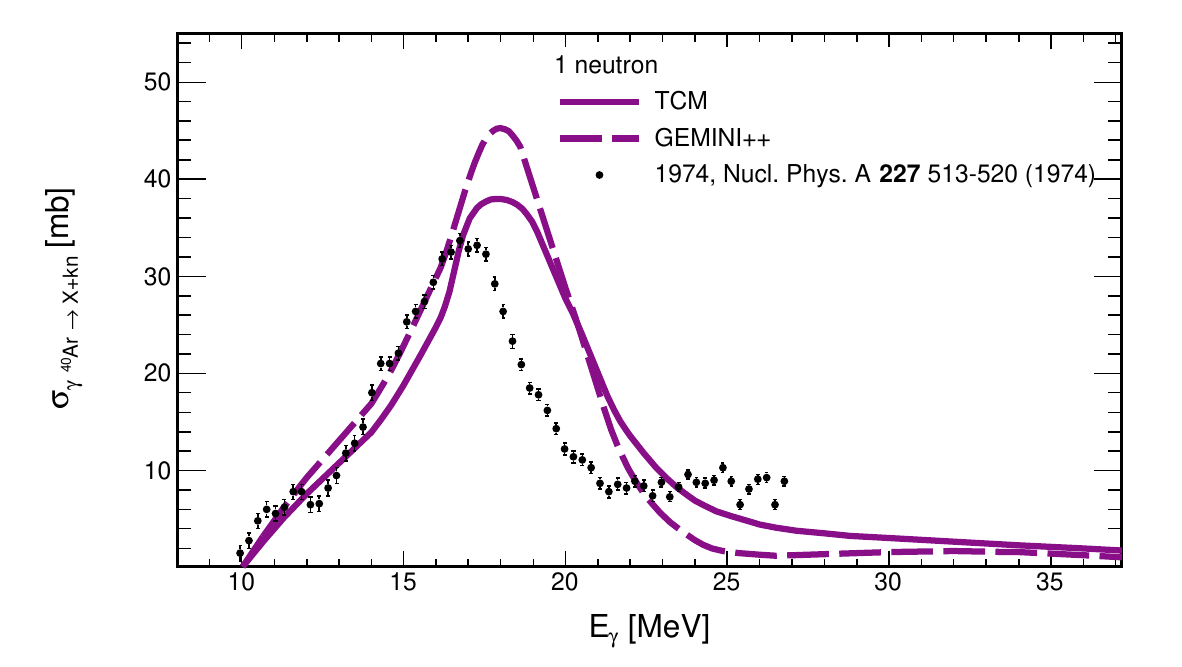}}
\put(-0.2,5.1){\bf \large a)}
	\put(-0.5,0){\includegraphics[width=0.5\textwidth]{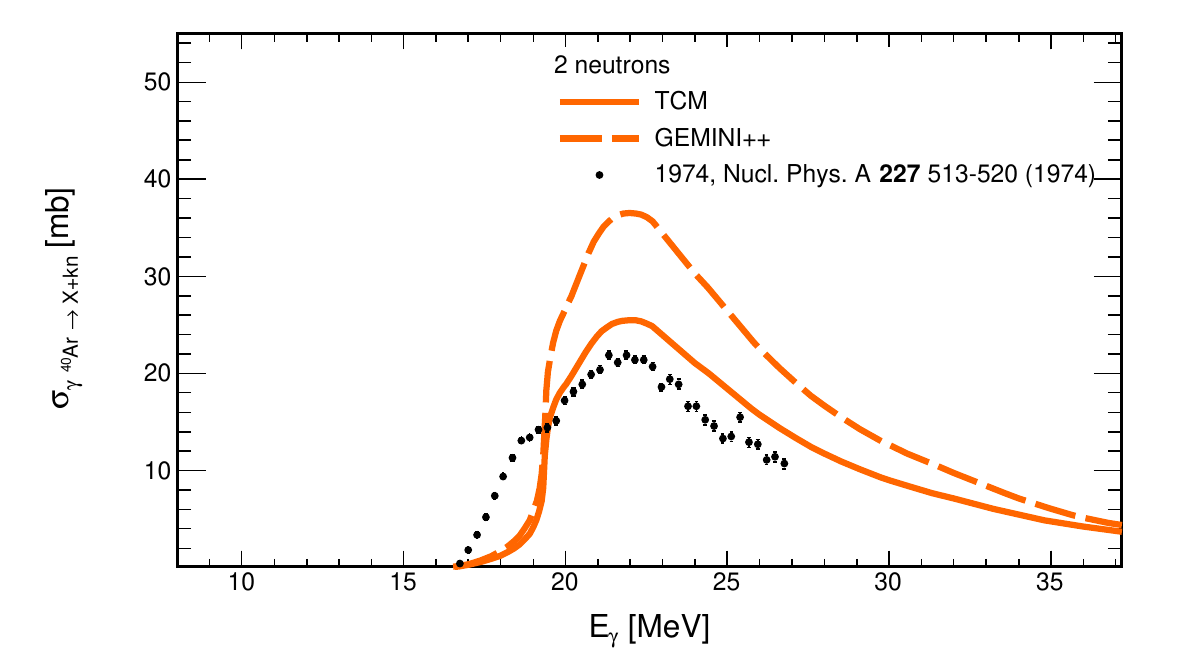}}
    \put(-0.2,2.3){\bf \large b)} 
    \end{picture}
	\caption{Cross section as a function of $E_{\gamma}$ for emission of a) 1, and b) 2 neutrons through the reaction $\gamma+^{40}\text{Ar}\to \text{X}+i\text{n}$ compared with experimental data from~\cite{VEYSSIERE_1974}.}
	\label{fig:neutron_cs_dist_exp}
\end{figure}

We assume that mesons (pions and kaons) are emitted during the first phase described by GiBUU. The reaggregation of nucleons into bound systems results in the formation of hot nuclei, which have to be de-exited. This part of the process is simulated by the Monte Carlo model GEMINI++ \cite{gemini}, where the particle and $\gamma$ emission, as well as the fission, are calculated in a consistent manner. 
The low- and intermediate-energy $\gamma$-nucleus collisions are instead treated with the Two Component Model (TCM) \cite{Jucha2025}.
This simple model has been created to mimic the fact that, for $E_{\gamma}>$~30~MeV, only a part of the photon energy is dissipated into the internal energy of the hit nucleus. This fact is taken into account within the TCM~\cite{Jucha2025}.

The comparison of low-energy photon spectra obtained with GEMINI++ and TCM with the experimental data \cite{VEYSSIERE_1974} is shown in Fig.~\ref{fig:neutron_cs_dist_exp}. The TCM provides a qualitative description of the 2 neutron emission but the centroid for 1 neutron emission is shifted towards higher energies.

\begin{figure}[!bt]
\setlength{\unitlength}{0.1\textwidth}
\begin{picture}(4,5.6)
\put(-0.5,2.6){\includegraphics[width=0.5\textwidth]{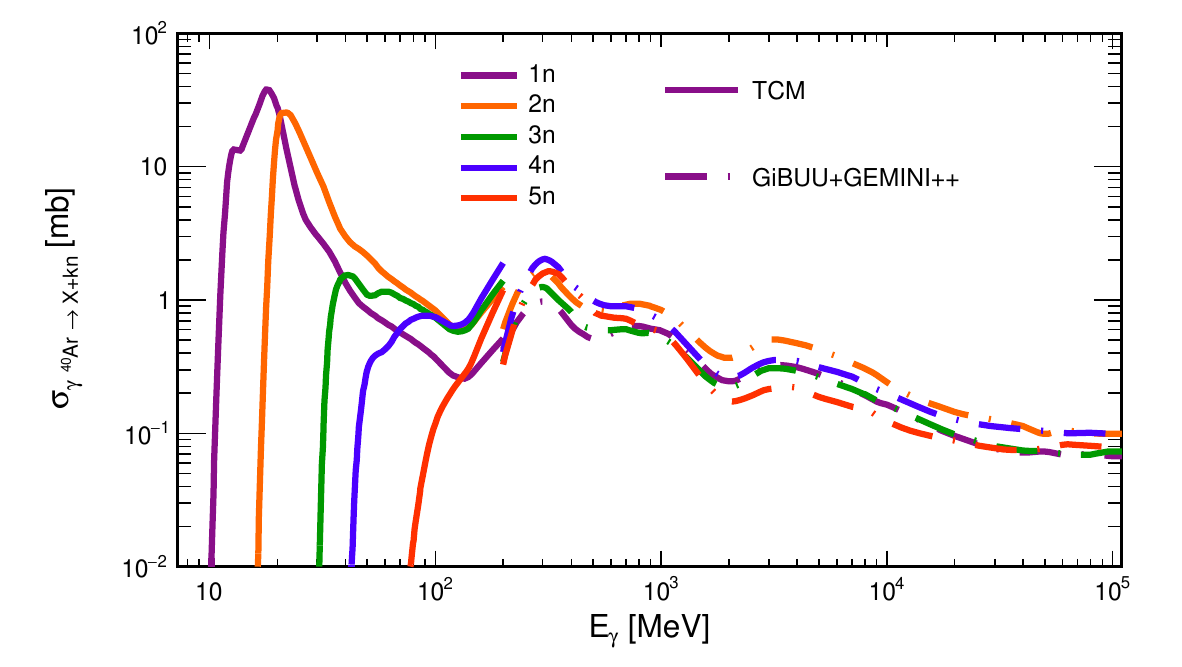}}/
\put(-0.2,4.9){\bf \large a)}
	\put(-0.5,0){\includegraphics[width=0.5\textwidth]{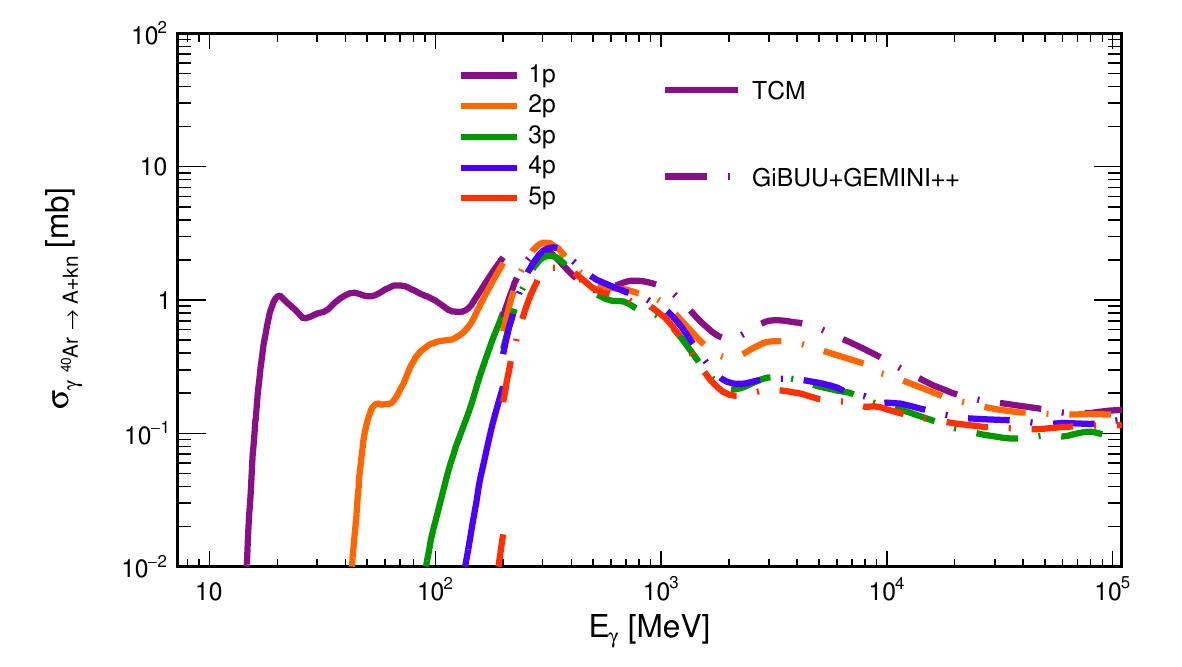}}
    \put(-0.2,2.3){\bf \large b)} 
    \end{picture}
	\caption{Cross section as a function of E$_{\gamma}$ for emission of 1-5 (a) neutrons (b) protons in  the reactions $\gamma$ + $^{40}$Ar$\to$X+\textit{i}n(p).}
	\label{fig:proton_cs_dist}
\end{figure}

The cross section for the emission
of 1, 2, 3, 4 and 5 neutrons (protons) is shown in Fig.~\ref{fig:proton_cs_dist}a (Fig.~\ref{fig:proton_cs_dist}b).

The emission of particles from the nucleus hit by photons at energy E$_{\gamma}$~$\leq$~200~MeV is obtained with the TCM, while for higher energies, the GiBUU+GEMINI++ hybrid method was applied. 
Similar results for $\gamma+^{208}\text{Pb}\to \text{X}+i\text{p}$ reaction can be found in \cite{Jucha2025, Jucha2026}. The emission of the nucleons stabilizes above E$_{\gamma}$~$\approx$~200~MeV as other channels start to play a dominant role. The production of mesons becomes more and more probable with increasing photon energy, which will be discussed in the next section.
\begin{figure}[bt]
	\centering
	\includegraphics[width=0.47\textwidth]{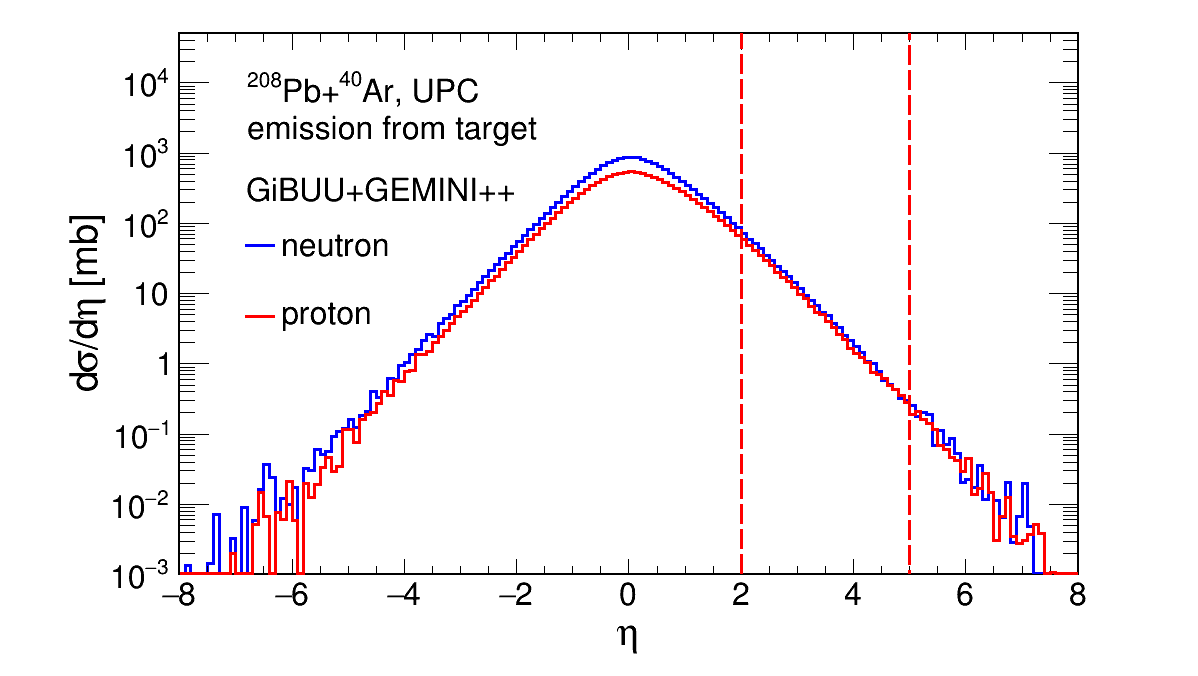}
	\caption{
    Estimated pseudorapidity distribution of nucleons emitted from excited $^{40}\text{Ar}$ target nuclei in UPC with $^{208}\text{Pb}$ beam nuclei. The blue line represents neutrons, the red line protons. The results refer to the full phase-space. The red dashed lines indicate the pseudorapidity range covered by the LHCb geometrical acceptance.
    }
	\label{fig:fullspace_UPC_arpb_nuc}
\end{figure}
\begin{figure}[!bt]
\setlength{\unitlength}{0.1\textwidth}
\begin{picture}(4,5.6)
\put(-0.5,2.8){\includegraphics[width=0.5\textwidth]{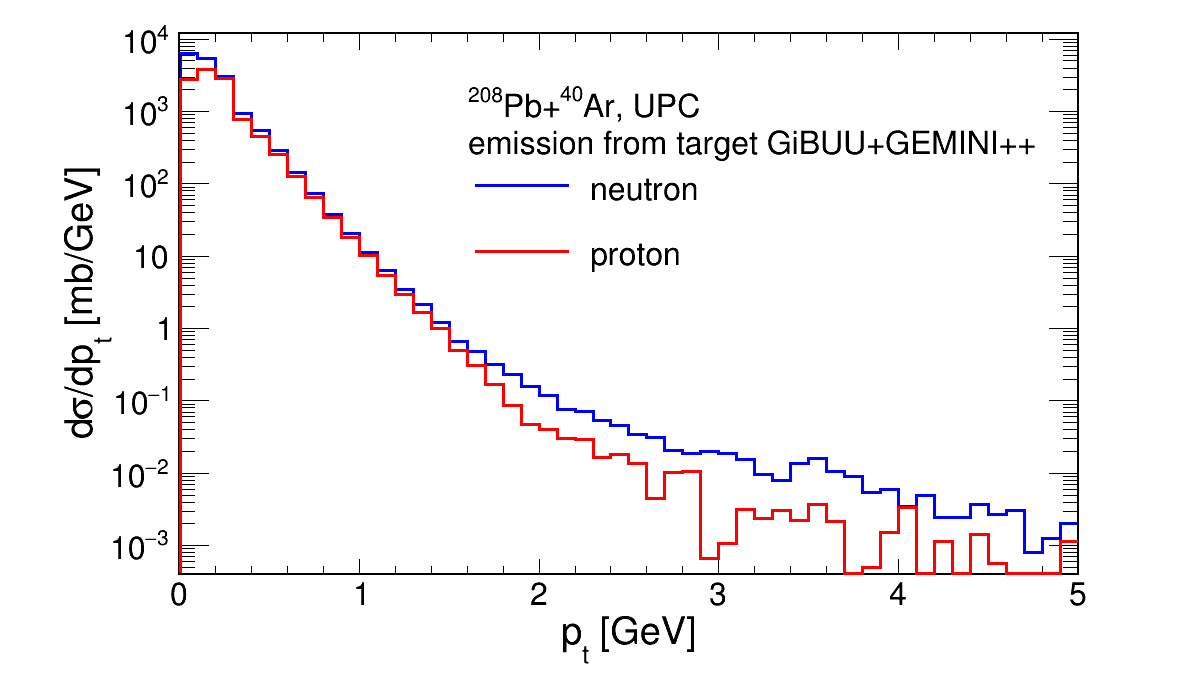}}
\put(-0.5,5.2){\bf \large a)}
	\put(-0.5,0){\includegraphics[width=0.5\textwidth]{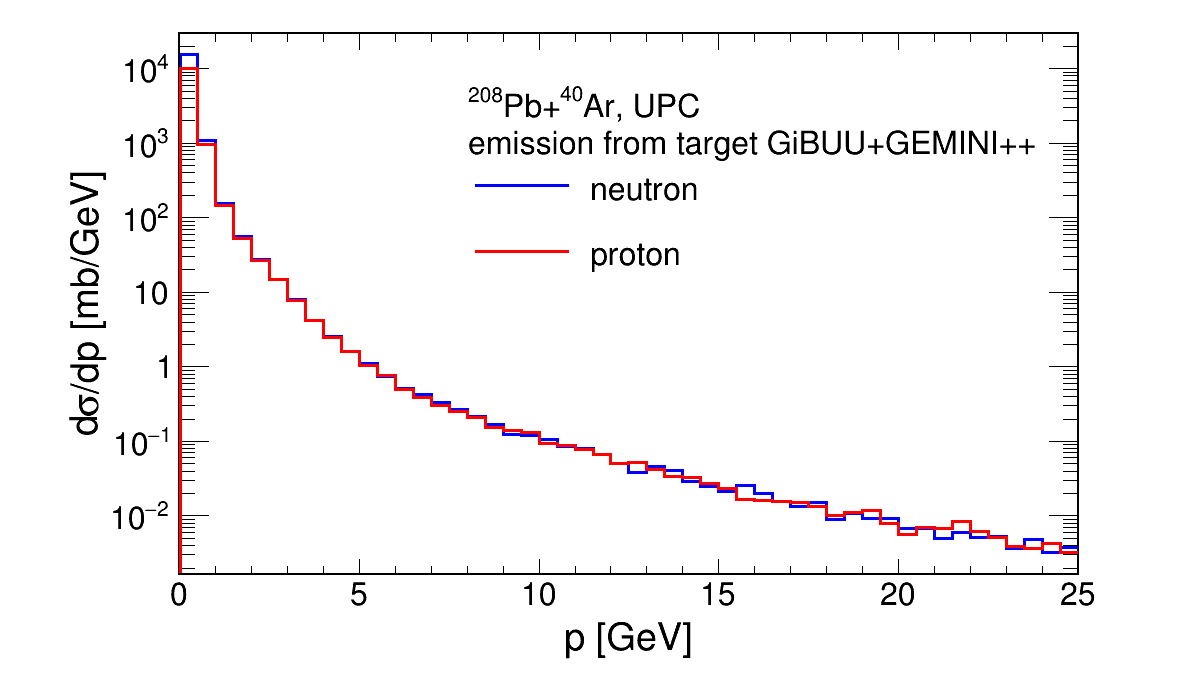}}
    \put(-0.5,2.3){\bf \large b)} 
    \end{picture}
	\caption{Differential cross section in a) transverse momentum and b) momentum for proton (red line) and neutron (blue line) emissions from a $^{40}\text{Ar}$ excited target. The results are obtained for the full phase-space.
    }
	\label{fig:fullspace_UPC_ar_nuc}
\end{figure}

The nucleons emitted from the excited target $^{40}\text{Ar}$ nucleus are almost symmetrical around pseudorapidity $\eta=0$, as shown in Fig.~\ref{fig:fullspace_UPC_arpb_nuc}. Most of these particles have a small pseudorapidity as there is no extra boost
(the emitting nucleus being at rest). The distribution of the nucleons emitted by the $^{208}$Pb beam nuclei is shifted, narrower and centered at the beam rapidity ($\eta_{beam}$ $>$ 10).
Such particles are emitted in very forward directions, outside the LHCb detector acceptance. The transverse momentum and total momentum
distributions for protons and neutrons are similar (Fig.~\ref{fig:fullspace_UPC_ar_nuc}).

\begin{table}[!bt]
\caption{
Absolute cross sections and relative uncertainties for multiple nucleon emission in fixed-target $^{208}\text{Pb}+^{40}\text{Ar}$  UPC (second column). The corresponding results for fixed-target $^{208}\text{Pb}+^{208}\text{Pb}$ UPC, with a hypothetical $^{208}\text{Pb}$  target, are also shown (third column) for comparison. Both results are obtained in the full phase-space with the TCM+GiBUU approach assuming a maximum photon energy of $E_{\gamma, max}$ = 70 GeV.
\\}
		\centering
	\begin{tabular}{lccc}
		\hline\\
	 	 &$\quad$ $^{208}$Pb+$^{40}$Ar [b] $\quad$   & $^{208}$Pb+$^{208}$Pb [b]$ \quad$   & $\sigma_{Pb}/\sigma_{Ar}$ \\\\
	     \hline\\
		1n & 4.23 $\pm$ 0.28  & 37.10 $\pm$ 2.81 & 8.8\\
		2n & 2.92 $\pm$ 0.34  & 9.60 $\pm$ 1.28  &3.3 \\
		3n & 0.61 $\pm$ 0.03  & 1.48 $\pm$ 0.16   &2.4 \\
		4n & 0.51 $\pm$ 0.03  & 1.19 $\pm$ 0.10 &2.3  \\
		5n & 0.36 $\pm$ 0.02  & 0.82 $\pm$ 0.06  &2.3 \\ \\\hline\\
		1p & 0.98 $\pm$ 0.04  & 1.36 $\pm$ 0.05  &1.4 \\
		2p & 0.56 $\pm$ 0.03   & 1.20 $\pm$ 0.01 & 2.1  \\
		3p & 0.43 $\pm$ 0.01   & 1.27 $\pm$ 0.01 &3.0  \\ 
        4p & 0.409 $\pm$ 0.004   & 1.239 $\pm$ 0.004  &3.0 \\ 
        5p & 0.328 $\pm$ 0.002   & 1.120 $\pm$ 0.004 &3.4 \\ \\\hline
	\end{tabular}
	\label{tab:results1}
\end{table}

The cross section for a given number of nucleons emission from the excited $^{40}\text{Ar}$ target nuclei and from a hypothetical $^{208}\text{Pb}$ target (considered here only for reference), are reported in Tab.~\ref{tab:results1} together with the corresponding uncertainties, calculated according to Eq.~\ref{deltas}. The single proton and single neutron emission from both targets is the most probable process, with an estimated relative uncertainty of less than 10\%.

The proton results shown in Tab.~\ref{tab:results1} require some additional comments.
They were obtained assuming the matching point
between application of TCM and 
GiBUU+GEMINI++ at $E_{\gamma}$ = 200 MeV.
While this matching point for $^{40}$Ar looks quite good (see Fig.~\ref{fig:proton_cs_dist}),
the matching point for $^{208}$Pb is more difficult to unambiguously 
identify. For instance in \cite{Jucha2026} it was taken somewhat larger. 
The results for auxiliary calculation for $^{208}$Pb depend rather strongly
 on the choice of the matching point. The fact that
the production cross section for three protons is larger than that for the production of
two protons (see Table~\ref{tab:results1}) is not true for 
the matching point $E_{\gamma} > $ 200 MeV.
In contrast, the results for $^{40}$Ar only weakly depend on the
choice of the matching point. Thus, we conclude that the results for $^{40}$Ar
are rather robust and can be verified by the LHCb collaboration.

The estimated uncertainties include the error estimation coming from low- and high-energy contributions as well as from the interpolation method
\begin{equation}
	(\Delta \sigma)^2 = (\Delta \sigma_\text{TCM})^2 + (\Delta \sigma_\text{GiBUU})^2 + (\Delta \sigma_\text{int})^2. \label{deltas}
\end{equation}

The low-energy part of the uncertainty comes from the free parameter $E_0$ of TCM. In our previous work~\cite{Jucha2025}
we discussed the impact of $E_0$ in a range from 20 MeV to 70 MeV. Here, we use these limiting values to estimate the uncertainty
\begin{equation}
	\Delta \sigma_{i,\text{TCM}} = \sigma(E_\gamma,i) \sqrt{ (\Delta P_{min})^2 + (\Delta P_{max})^2 },
\end{equation}
where $\Delta P_{min}=|P_{E_0=50\text{ MeV}} - P_{E_0=20\text{ MeV}}|$ and $\Delta P_{max}=|P_{E_0=50\text{ MeV}} - P_{E_0=70\text{ MeV}}|$ are the relative differences between the probability obtained with three different free TCM parameters: $E_0$ = 50 MeV, which is the optimal value used in previous articles, and $E_0$ = 20 MeV and $E_0$ = 70 MeV which correspond to the minimal and maximal value of this parameter. The total uncertainty from TCM is estimated as
\begin{eqnarray}
		\Delta \sigma_\text{TCM} = \sqrt{\sum_i n_i^2 \Delta \sigma_{i,\text{TCM}}^2 \Delta E_\gamma^2 },
\end{eqnarray}
where $n_i = \int d^2b N(E_{\gamma,i},b) dE_{\gamma,i}$ is the photon flux at a given photon energy. Here, the index $i$ denotes the $i$-th photon-energy point.

Similarly, the uncertainty coming from the high-energy photon part, calculated with the GiBUU model is obtained as
\begin{equation}
    \Delta \sigma_\text{GiBUU} = \sqrt{\sum_i n_i^2 \Delta \sigma_{i,\text{GiBUU}}^2 \Delta E_\gamma^2 }.
\end{equation}
The $\Delta \sigma_{i,\text{GiBUU}}$ is the statistical uncertainty of the GiBUU simulation in the $i$-th set of data
calculated for a given photon energy
\begin{eqnarray}   
\Delta \sigma_{i,\text{GiBUU}} = \sqrt{ \sum^{N_\text{event}}_{j=1} \bigg(\sigma_j \bigg)^2 }.
\end{eqnarray}

The last source of uncertainty is the interpolation between data sets. As mentioned before, the cross section
from TCM and GiBUU is calculated for discrete values of the photon energy.
The connection between adjacent points is determined by a linear interpolation method.
To compute the uncertainty we also applied a different method utilizing the Bézier curve \cite{farin2002curves} with smoothstep function:
\begin{equation}
  y(x) =  y(x_1) + (y_2-y_1) \cdot t^2(3-2t) , \hspace{0.2cm} t = \frac{x-x_1}{x_2-x_1},
\end{equation}

\noindent where $y(x)$ is the interpolated value, and $y(x_1), y(x_2)$ are known values of the function in point $x_1, x_2$ respectively. The difference between the total cross sections calculated with both methods provides the final uncertainty estimate:
\begin{equation}
	\Delta \sigma_\text{int} = |\sigma_\text{linear} -  \sigma_\text{Bézier}|.
\end{equation}

\section{Meson Emission}
\label{Sect.IV}
The GiBUU approach includes many different processes, important in high-energy heavy-ions collisions. It is also well known to reproduce the experimental data for meson production for $E_\gamma$~$<$~1~GeV. 

\begin{figure}[!bt]
	\centering
	\includegraphics[width=0.5\textwidth]{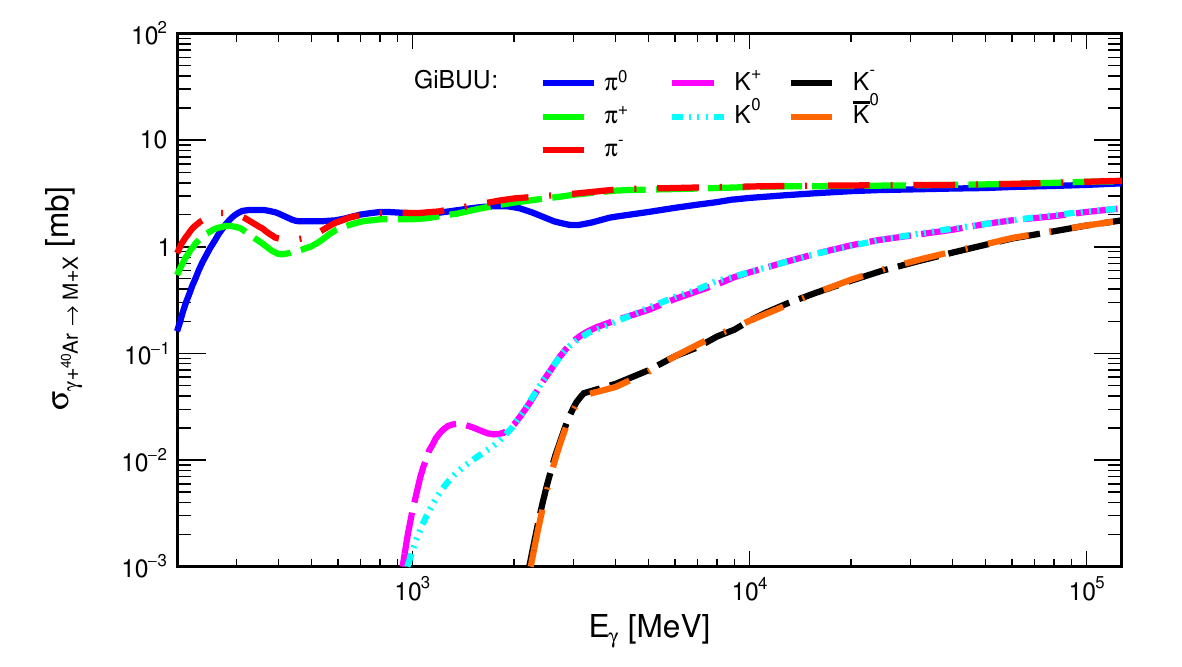}
	\caption{Cross section for emission of mesons in the reaction $\gamma$+$^{40}$Ar$\to$ M+X as a function of $E_\gamma$. }
	\label{fig:one_part_dist}
\end{figure}

The photoproduction cross section for charged and neutral pions and kaons is shown in Fig.~\ref{fig:one_part_dist} 
as a function of the photon energy. As expected, pions can be produced at a much lower
excitation energy (photon energy) than kaons as their masses are smaller and different phenomena are responsible for their creation. Furthermore, the energy threshold demanded for associated strangeness production decreases the kaon production probabilities compared to the pion production.

The GiBUU pion production in $\gamma$-induced reactions has been implemented based
on the available experimental data. Dedicated experiments for meson photoproduction
have been performed at the Mainz Microtron (MAMI) facility, at Frascati INFN Laboratories (LNF), at ELSA
in Bonn and at Hall B of the Jefferson Laboratory
(for a review of the main experiments in this field see e.g. \cite{BIANCHI1994333,PhysRevC.54.1688,PhysRevC.60.064616}).
Both resonant and non-resonant mechanisms of pion production in $\gamma$-A reactions are considered. The dominant contribution comes from the $\Delta$-resonance region.

For the calculation of the kaon photoproduction in GiBUU, in general, the coupled-channels formalism is applied. It makes use of elementary photo-production data measured, e.g. by the LEPS, CLAS, SAPHIR or CBELSA collaborations.
For a review of experimental data and the description of the model, see \cite{PhysRevC.88.055204}.
Also here, initial excitation of resonances plays the dominant role, but non-resonant production is considered as well.

\begin{table}[!bt]
	\caption{
    Nucleons and mesons emission cross section from the $^{40}\text{Ar}$ target (second column) and from the hypothetical $^{208}\text{Pb}$ target (third column) in UPC with the $^{208}\text{Pb}$ LHC beam. The results are obtained in the full phase-space with the GiBUU approach using $E_{\gamma, max}$ = 70 GeV.
    \\}
	\centering
	\begin{tabular}{lrrc}
		\hline\\
		 & $^{208}$Pb+$^{40}$Ar [mb] &$\quad$ $^{208}$Pb+$^{208}$Pb [mb] $\quad$ & $\sigma_{Pb}/\sigma_{Ar}$ \\\\
		\hline\\
		p			  & 3635 $\pm$ 8  	  & 10927 $\pm$ 66    $\quad$   & 3.0 \\
		n			  & 10475 $\pm$ 6  	  & 65328 $\pm$ 59   $\quad$    & 6.2 \\&&&\\
		$\pi^0$ 	  & 948 $\pm$ 2	  & 2603 $\pm$ 11  $\quad$& 2.8 \\
		$\pi^+$ 	  & 1010 $\pm$ 3  & 2496 $\pm$ 10 $\quad$ & 2.5 \\
		$\pi^-$ 	  & 1123 $\pm$ 3  & 3335 $\pm$ 13 $\quad$ & 3.0 \\&&&\\
		K$^+$ 	  		  & 73.5 $\pm$ 0.6  & 214.4 $\pm$ 1.8   $\quad$ & 2.9 \\
		K$^0$ 		  & 74.1 $\pm$ 0.6  & 229.0 $\pm$ 1.9  $\quad$  & 3.1 \\
		K$^-$ 	  & 30.7 $\pm$ 0.4  & 61.5 $\pm$ 0.8	 $\quad$& 2.0 \\
		$\bar{\text{K}}^0$   & 31.0 $\pm$ 0.4  & 61.3 $\pm$ 0.8	 $\quad$& 2.0 \\ \\\hline
	\end{tabular}
	\label{tab:results2}
\end{table}

\begin{figure*}[!bt]
\setlength{\unitlength}{0.1\textwidth}
\begin{picture}(9,9)
\put(-0.5,6){\includegraphics[width=0.54\textwidth]{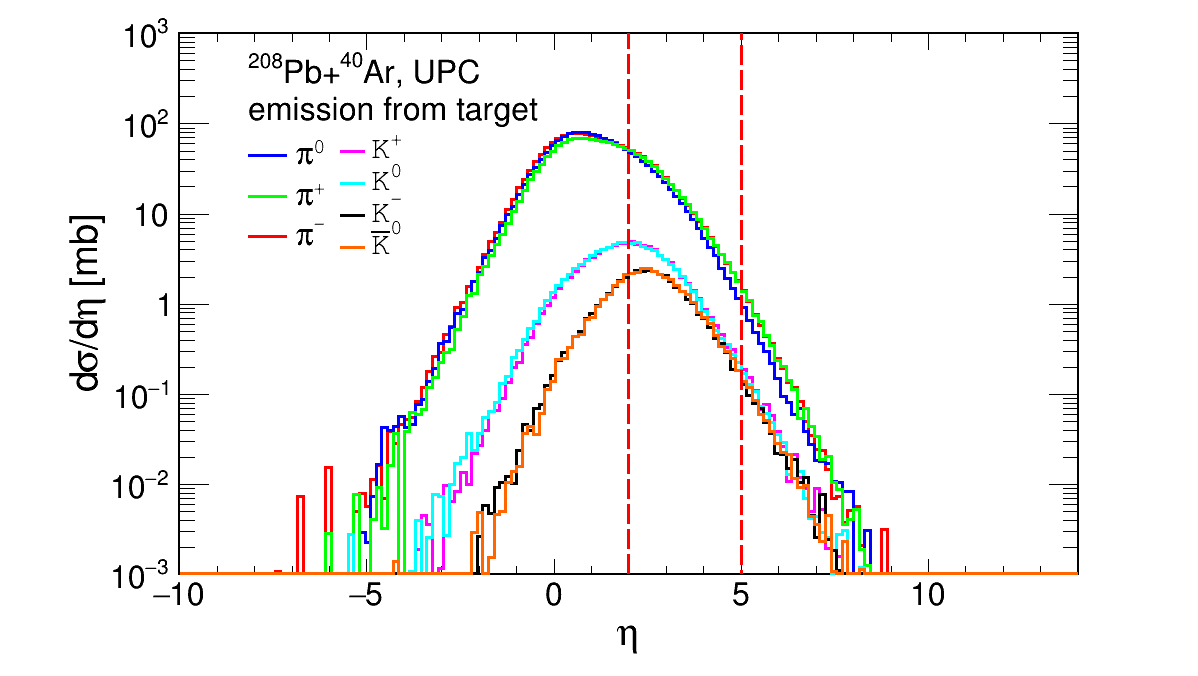}}
\put(-0.2,8.7){\bf \large a)}
\put(4.5,6){\includegraphics[width=0.54\textwidth]{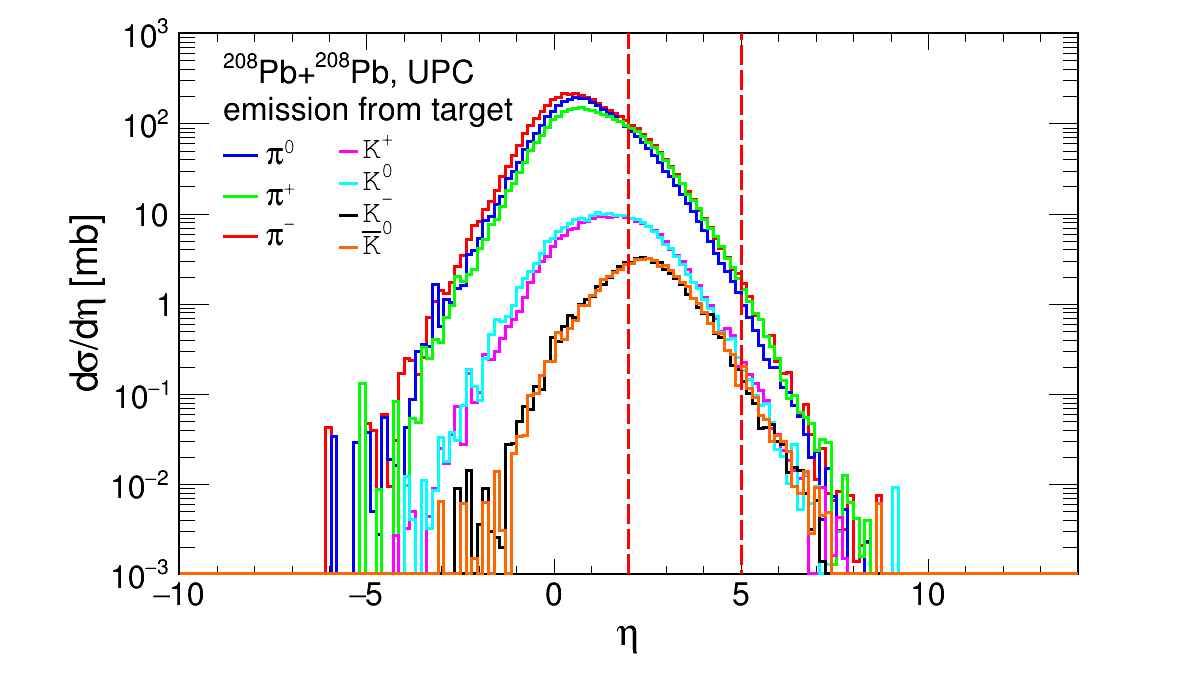}}
\put(4.7,8.7){\bf \large b)}

\put(-0.5,3){\includegraphics[width=0.54\textwidth]{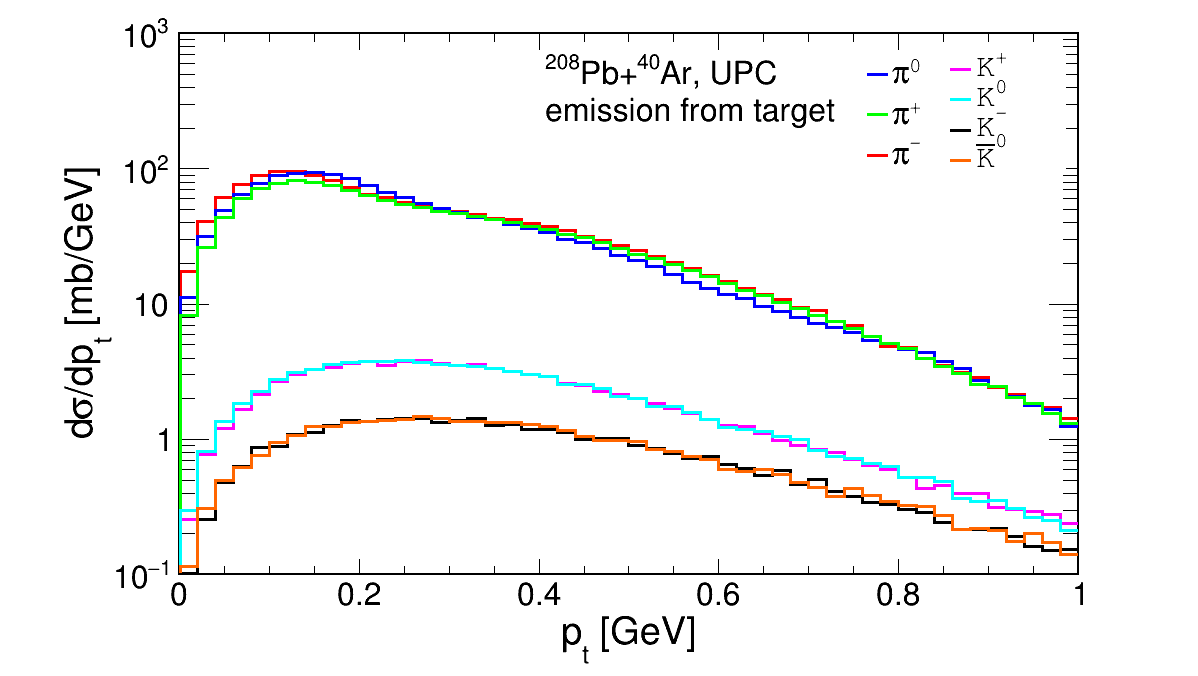}}
\put(-0.2,5.7){\bf \large c)}
\put(4.5,3){\includegraphics[width=0.54\textwidth]{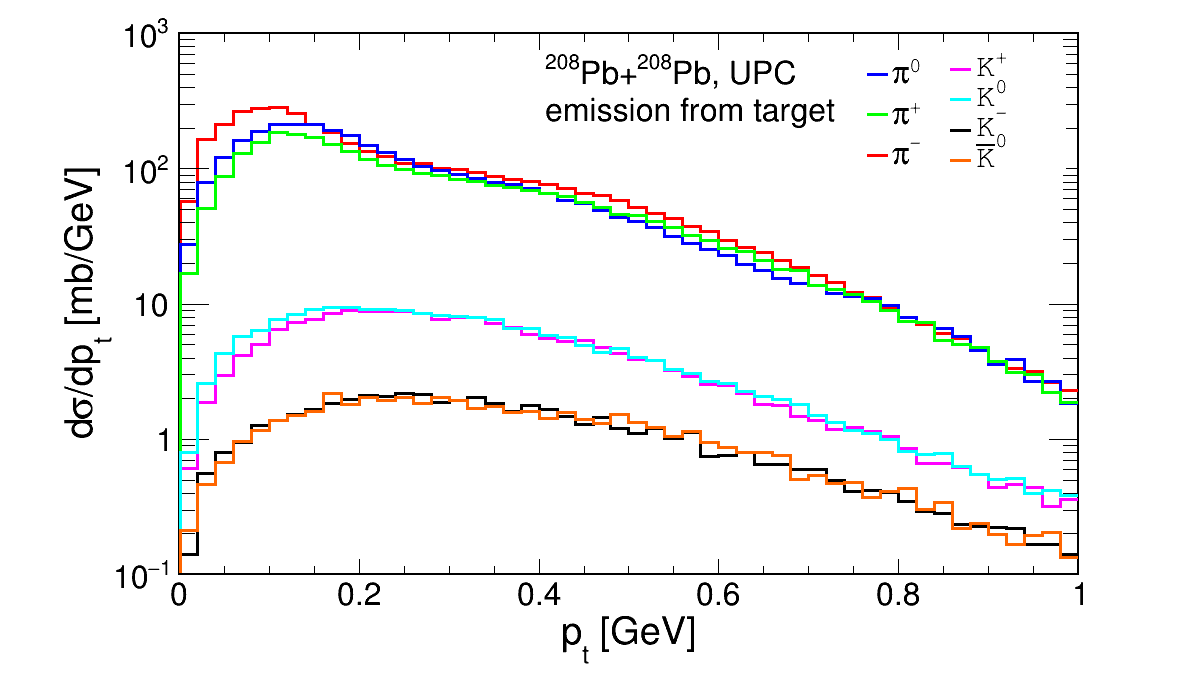}}
\put(4.7,5.7){\bf \large d)}
	\put(-0.5,0){\includegraphics[width=0.54\textwidth]{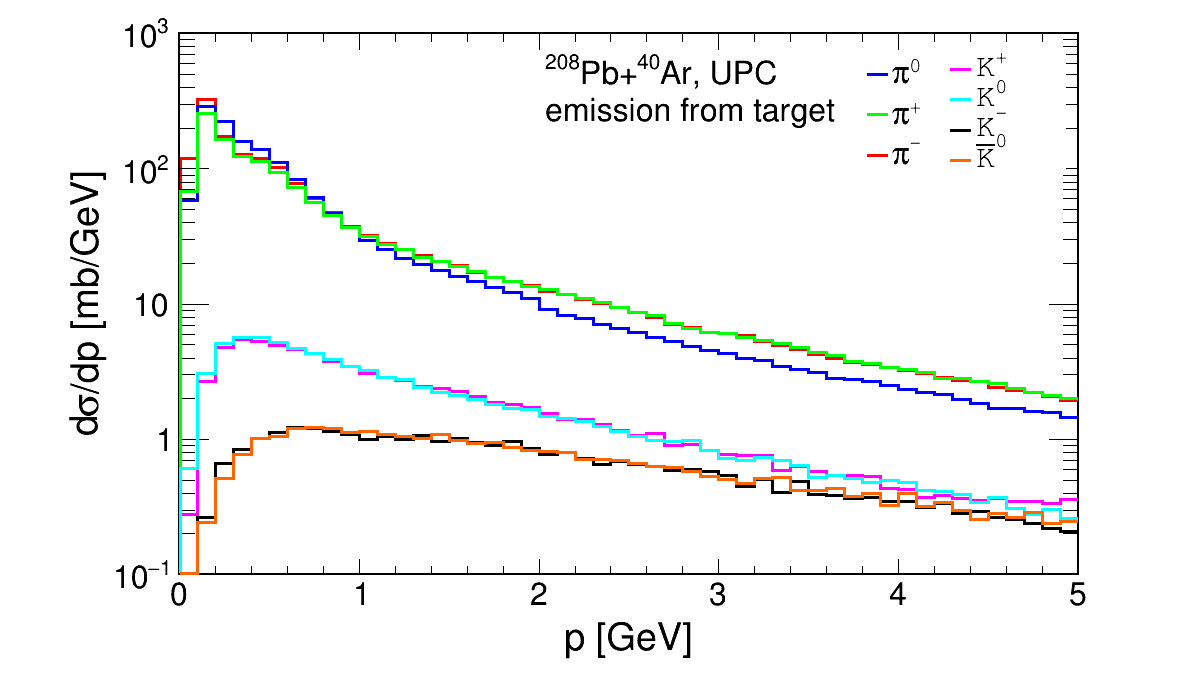}}
    \put(-0.2,2.7){\bf \large e)} 
	\put(4.5,0){\includegraphics[width=0.54\textwidth]{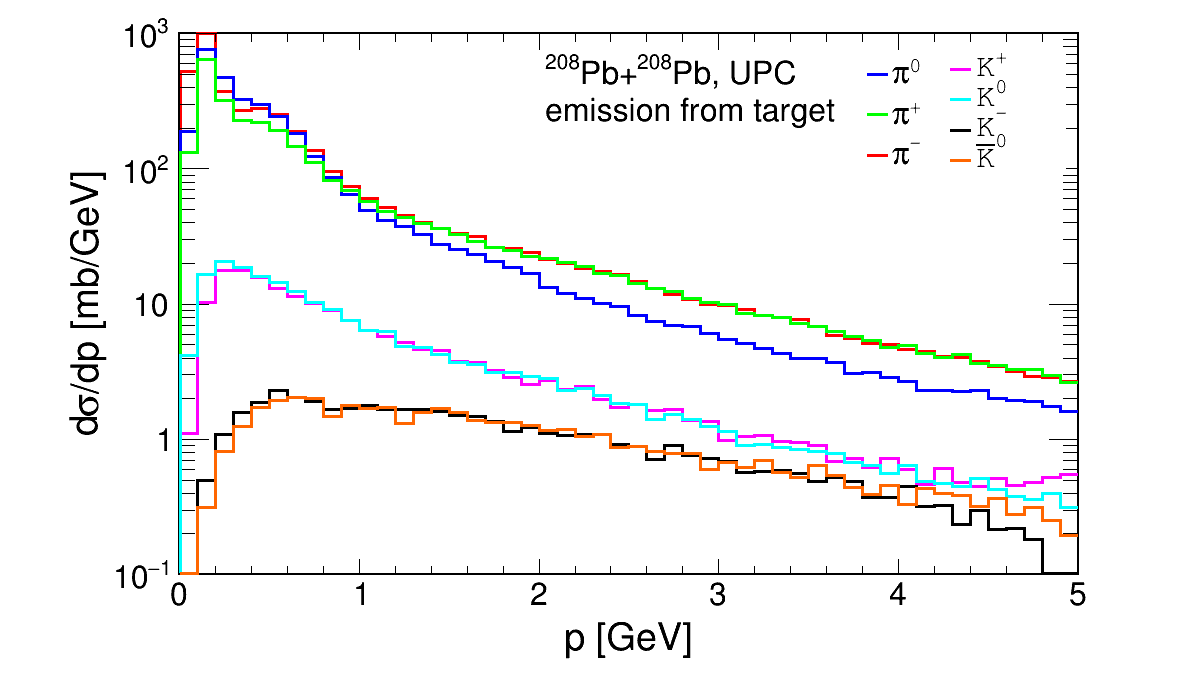}}
    \put(4.7,2.7){\bf \large f)} 
    \end{picture}
	\caption{Distributions of mesons from $^{208}\text{Pb}$+$^{40}\text{Ar}$ (left) and $^{208}\text{Pb}$+$^{208}\text{Pb}$ (right) fixed-target collisions in a,b) pseudorapidity, c,d) transverse momentum,  e,f) momentum. The blue line represents $\pi^0$, the green line $\pi^+$, the red line $\pi^-$, the magenta line K$^+$, the cyan line K$^0$, the black line K$^-$, and the orange line $\bar{\text{K}}^0$. The results are obtained in the full phase-space.}
	\label{fig:fullspace_UPC_pb}
\end{figure*}

Both pions and kaons, when produced, are propagated through the medium of the bombarded nucleus, where secondary interactions with the nucleons are possible. This may affect their energies and diminish their emission probabilities.

Detailed considerations of an implementation of the $\gamma$-induced reaction in GiBUU and further links can be found in \cite{BUSS20121}.

Nucleons and mesons emission cross sections are collected in Tab.~\ref{tab:results2} for the $^{40}$Ar target and, for comparison, for a hypothetical $^{208}\text{Pb}$ target, both excited by incoming $^{208}\text{Pb}$ nuclei with  $E_{beam}/A$ = 2.51~TeV.

The production cross section for pions is slightly smaller than that for nucleons and depends on the mass number of the target nucleus. 
With the exception of the neutron case, the ratio of cross sections $\sigma_\text{Pb}/\sigma_\text{Ar}$ is around 3. A charge dependence is also observed (see Tab.~\ref{tab:results2}), with the $\pi^+$ ($K^+$) ratio being smaller (larger) than the corresponding ratio for negative pions (kaons).

The distributions of the pions and kaons in pseudorapidity, transverse momentum, and momentum are displayed in Fig.~\ref{fig:fullspace_UPC_pb}. In contrast to nucleons emission (Fig.~\ref{fig:fullspace_UPC_ar_nuc}), the pseudorapidity distribution of mesons is not centered at $\eta$~=~0, due to the specific non-equilibrium processes behind their production. Nucleons, coming from the Hauser-Feshbach evaporation, are independent of the direction of the photon momentum. The shapes of the transverse momentum and momentum distributions are found to be similar for $^{40}$Ar and $^{208}\text{Pb}$ targets.

\section{Experimental conditions}
\label{Sect.V}

Among the experiments running at the Large Hadron Collider, LHCb is presently the only one that can also be operated in fixed-target mode. The LHCb detector~\cite{LHCb_1,LHCb_2} is a general-purpose forward spectrometer specialized in detecting hadrons containing $c$ and $b$ quarks. It is fully instrumented in the $2<\eta<5$ region with state-of-the-art particle detectors including a Vertex Locator (VELO), a tracking system, two Cherenkov (RICH) detectors, electromagnetic and hadronic calorimeters and a muon detector, ensuring excellent vertexing, tracking and particle identification capabilities.

\begin{figure}[!bt]
\setlength{\unitlength}{0.1\textwidth}
\begin{picture}(4,5.6)
\put(-0.5,2.8){\includegraphics[width=0.5\textwidth]{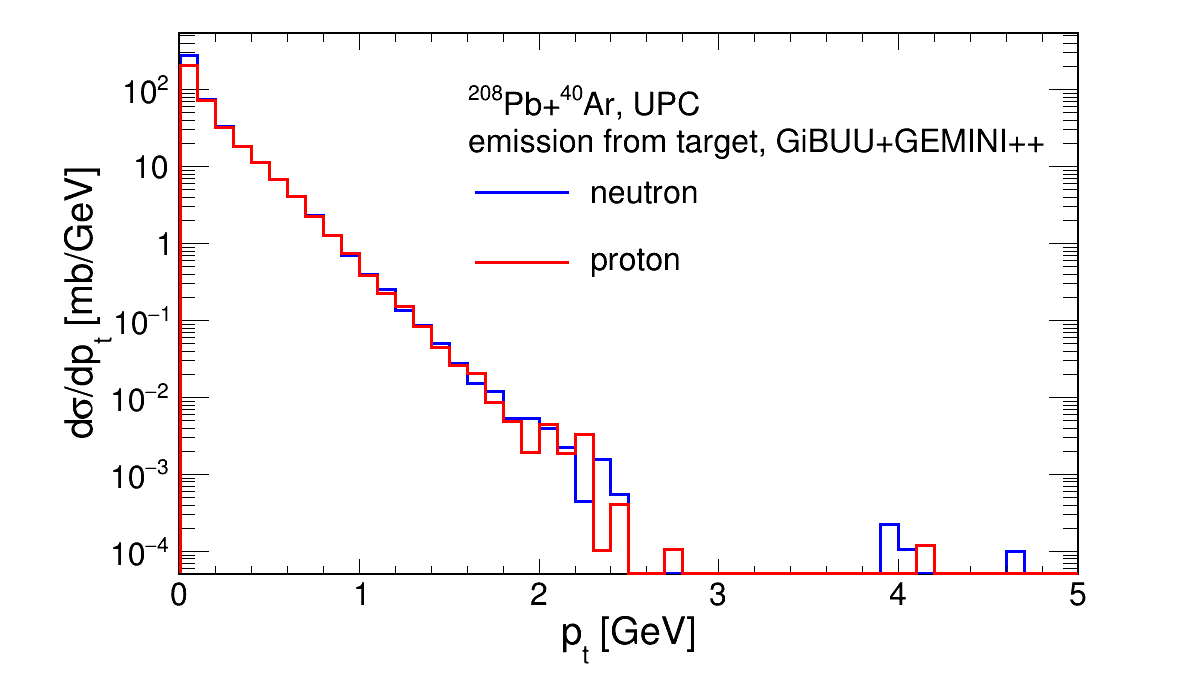}}
\put(-0.3,5.2){\bf \large a)}
	\put(-0.5,0){\includegraphics[width=0.5\textwidth]{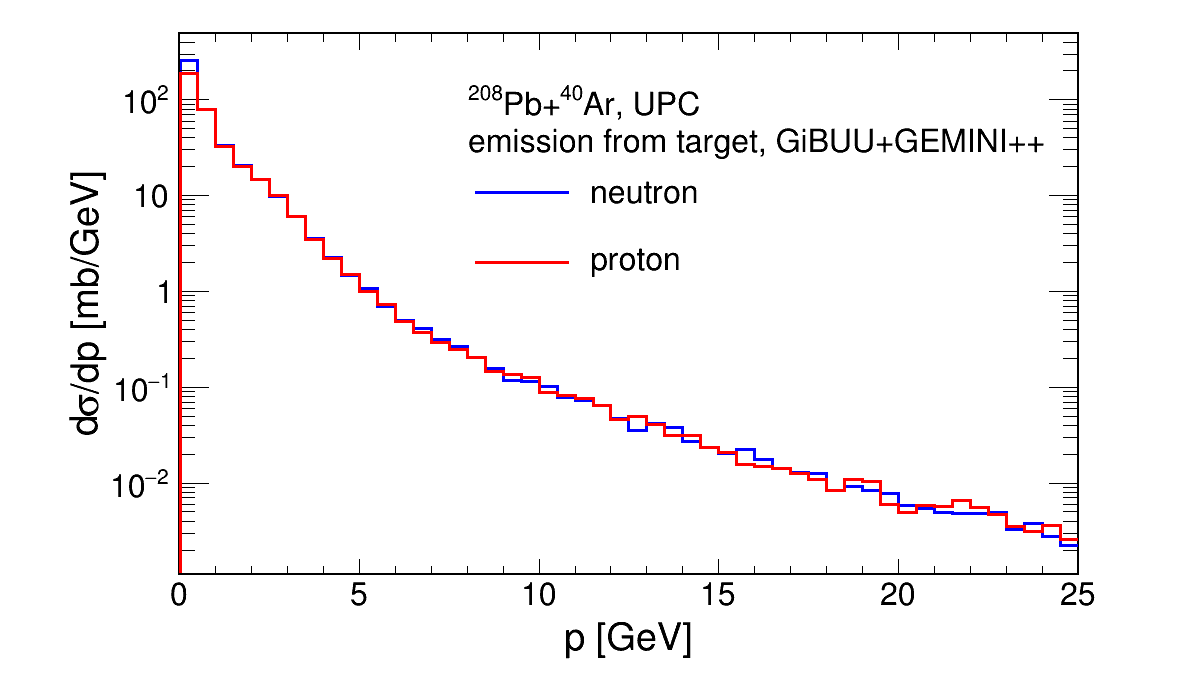}}
    \put(-0.3,2.3){\bf \large b)} 
    \end{picture}
	\caption{Similar to Fig.~\ref{fig:fullspace_UPC_ar_nuc} but restricted to the LHCb pseudorapidity range $2<\eta<5$.}
	\label{fig:LHCb_UPC_ar_nuc}
\end{figure}

Fixed-target collisions at LHCb have been studied since the beginning of Run 2 (2015), exploiting the SMOG system~\cite{SMOG}, which enables the injection of low-pressure noble gasses into the beam pipe section that crosses the VELO. In particular, p-He, p-Ne, p-Ar as well as Pb-Ne and Pb-Ar collisions have been performed at $\sqrt{s_{NN}} \approx 70-110$ GeV, paving the way to a diverse physics program ranging from the nucleon structure to heavy-ion physics as well as cosmic rays physics and dark matter search, at unique kinematic conditions.

The SMOG system has been recently upgraded with the installation, in front of the VELO, of an openable storage cell for the target gas, which ensures target areal densities higher by up to two orders of magnitude compared to SMOG. The upgraded system, SMOG2~\cite{SMOG2}, successfully commissioned at the beginning of Run 3 (2022), also comprises a new and more sophisticated Gas Feed System which allows to precisely measure the injected gas density and to inject several more gas species, including $\rm{H_2}$, $\rm{D_2}$ and Xe. Thanks to the well separated beam-beam and beam-gas interaction regions, achievable with the use of the storage cell, simultaneous collider and fixed-target data taking are possible and routinely performed at LHCb.

Particles produced in fixed-target collisions within the LHCb acceptance are collected and identified by the LHCb spectrometer. Specifically, protons/antiprotons as well as charged pions and kaons are efficiently reconstructed by the tracking system and cleanly discriminated by the RICH detectors in a wide momentum range, whereas neutral pions are reconstructed in the EM calorimeter exploiting their decay into photon pairs. To our knowledge, no specific PID tools have been developed so far for the identification of neutrons. 

Unlike ordinary ultrarelativistic heavy-ion collisions, in which hundreds of particles are typically produced in a single event, in ultraperipheral collisions the colliding ions remain intact and only a few particles are produced,
either exclusively, e.g. through pomeron exchange, or by evaporation of the excited colliding systems.
Thus, the main criterion to select experimentally such events requires a very low track multiplicity in the detector, especially in the forward/backward regions. At LHCb, this requirement of low rate in the detector is implemented by selecting specific trigger and stripping lines.
At the trigger level one requires a limited number ($<$ 50) of hits in the Scintillating Pad Detector (SPD)\footnote{The SPD, which was primarily used for the Level-0 trigger, has been removed during the Long Shutdown 2 as LHCb transitioned to a full all-software trigger system that reads out the entire detector event-by-event, eliminating the need for fast hardware-level pre-decisions.}, which until the LHC Run 2 was a part of the LHCb calorimetric system, low rate in the VELO pileup modules and at least 1 “VELO track”, i.e., a short track leaving signals only in the VELO detector. Then, a specific stripping line is used to select events with at least 1 “long track”, i.e., crossing the entire tracking system, and a limited number ($\le$ 10) of “VELO tracks”. Additional and stricter selection criteria on the event multiplicity (e.g. number of SPD hits $<$ 20, number of “long tracks” $\leq$ 2, etc.) are then typically applied in the data analyses to further improve the selection of UPC events. In particular, for Pb-Pb data collected by LHCb during the Run 2 of the LHC, a further constraint is imposed by requiring low activity in the HeRSCheL (High Rapidity Shower Counters for LHCb) detector\footnote{The HeRSCheL detector consists of five stations, two of which were positioned upstream (at z=20m and z=114m) and three downstream (at z=-7.5m, z=-20m and z=-114m) of the LHCb spectrometer. Each station is segmented into four scintillator plates read out by separate PMTs. The signal in the HeRSCheL stations is dominated by energy deposits from low-energy electrons and positrons produced in the showers induced by high-rapidity particles interacting with machine elements close to the scintillators.
The HeRSCheL detector has been decommissioned during the LHC Long Shutdown 2 due to radiation damage of the phototubes. The idea of installing a new and reduced version of the detector (the LASARUS project) is being discussed among the Collaboration.}, a forward/backward detector positioned upstream and downstream of the LHCb spectrometer at pseudorapidities  $5<|\eta|<10$.

\begin{table}[!bt]
		\caption{Absolute cross sections for multiple nucleon emission in fixed-target $^{208}\text{Pb}+^{40}\text{Ar}$ UPC (second column). The corresponding results for fixed-target $^{208}\text{Pb}+^{208}\text{Pb}$ UPC, with a hypothetical $^{208}\text{Pb}$ target, are reported in the third column for comparison. Both results are obtained within the LHCb acceptance (2~$<$~$\eta$~$<$~5) with the TCM+GiBUU approach assuming a maximum photon energy of $E_{\gamma, max}$ =~70~GeV.\\}
\centering
	\begin{tabular}{lccc}
		\hline\\
		 &$\quad$ $^{208}$Pb+$^{40}$Ar [mb]  & $\quad$$^{208}$Pb+$^{208}$Pb [mb] $\quad$  & $\sigma_{Pb}/\sigma_{Ar}$ \\ \\\hline\\ 
		1n & 473.42 $\pm$ 2.40  &  2711.4 $\pm$ 23.4 & 5.7 \\
		2n & 56.07 $\pm$ 0.59   & 510.7 $\pm$ 6.4 & 9.1 \\
		3n & 8.46 $\pm$ 0.13   & 91.2 $\pm$ 1.9   & 10.8 \\
		4n & 1.40 $\pm$ 0.04   & 14.5 $\pm$ 0.5   & 10.4 \\
		5n & 0.23 $\pm$ 0.01  & 4.7 $\pm$ 0.4    & 20.4 \\ \\\hline\\
		
		1p & 400.55 $\pm$ 1.58   & 1138.8 $\pm$ 12.0  & 2.8 \\
		2p & 41.26 $\pm$ 0.34   & 94.0 $\pm$ 2.0   & 2.3 \\
		3p & 6.33 $\pm$ 0.10   & 12.6 $\pm$ 0.7    & 2.0 \\
        4p & 1.09 $\pm$  0.03   & 2.35 $\pm$ 0.21  & 2.2 \\
        5p & 0.17 $\pm$ 0.01   & 0.49 $\pm$ 0.09   & 2.9 \\\\\hline
	\end{tabular}
	\label{tab:results_cut}
\end{table}

To match the effective experimental conditions, our predictions, provided in the full phase-space, have, in principle, to be convoluted with the geometrical and kinematic restrictions dictated by the detector features (acceptance, resolution, tracking and  PID efficiency, etc.) and the detection conditions. This is a very complicated task, which goes beyond the scope of this article. In the present work, we consider only the effects of the restricted geometrical acceptance of the LHCb detector and we simply quote some minimal cuts that are usually applied in LHCb data analyses. In addition to the geometrical acceptance constraint of $2<\eta<5$, off-line requirements on kinematic variables such as total momentum ($p$) and transverse momentum ($p_t$) of the detected particles are applied in the analyses of LHCb data. For instance, $p>$ 3 GeV and $p_t>$ 400 MeV ($p_t>$ 800 MeV) for charged pions and kaons (neutral pions) are minimal requirements in typical LHCb data analyses of fixed-target collisions, although additional and more strict requirements are usually applied, depending on the analysis subject and strategy.

The acceptance of the LHCb detector limits the possibility to detect very forward particles ($\eta > 5$) emitted by the excited beam nuclei, which can only be detected, e.g., with a Zero Degree Calorimeter, such as the one of the ALICE experiment \cite{ALICE:2012aa, ALICE502a, ALICE_protons}. A similar detector is, however, not available at LHCb. For this reason, in the present work, we consider only the particles emitted by the target nuclei. Due to their smaller average pseudorapidity, the particles (protons, mesons) easily fall within the acceptance of the LHCb spectrometer, thereby being detected and identified.

\begin{figure*}[!bt]
\setlength{\unitlength}{0.1\textwidth}
\setlength{\unitlength}{0.1\textwidth}
\begin{picture}(9,9)
\put(-0.5,6){\includegraphics[width=0.54\textwidth]{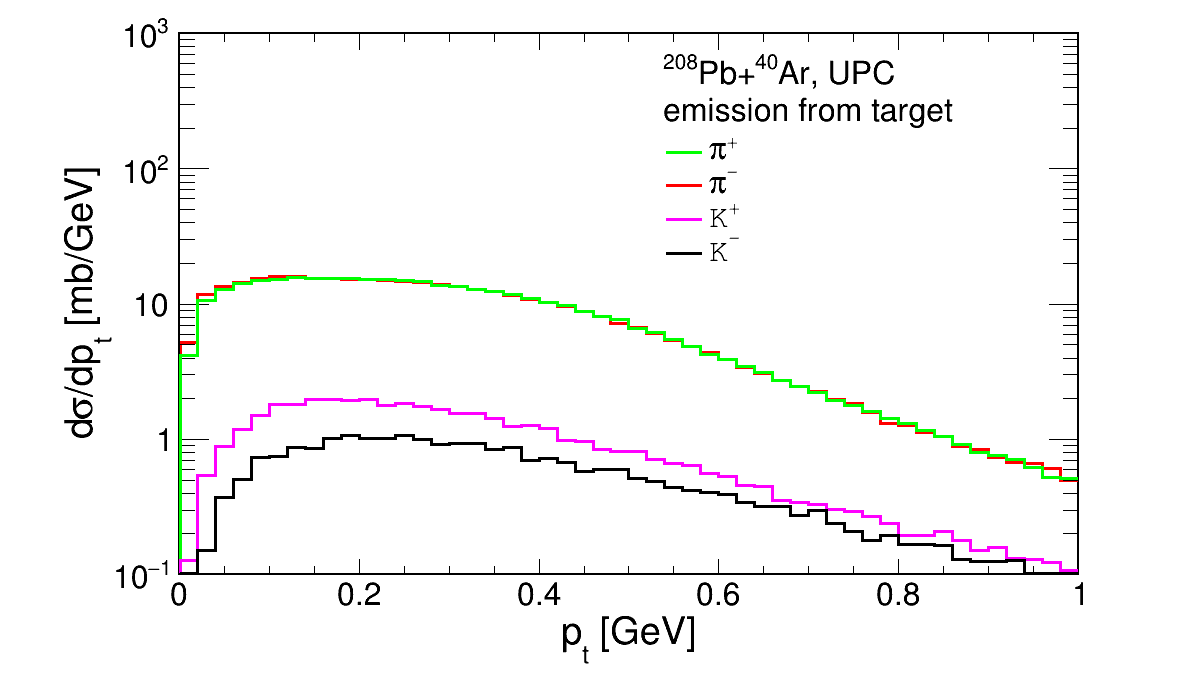}}
\put(-0.2,8.7){\bf \large a)}
\put(4.5,6){\includegraphics[width=0.54\textwidth]{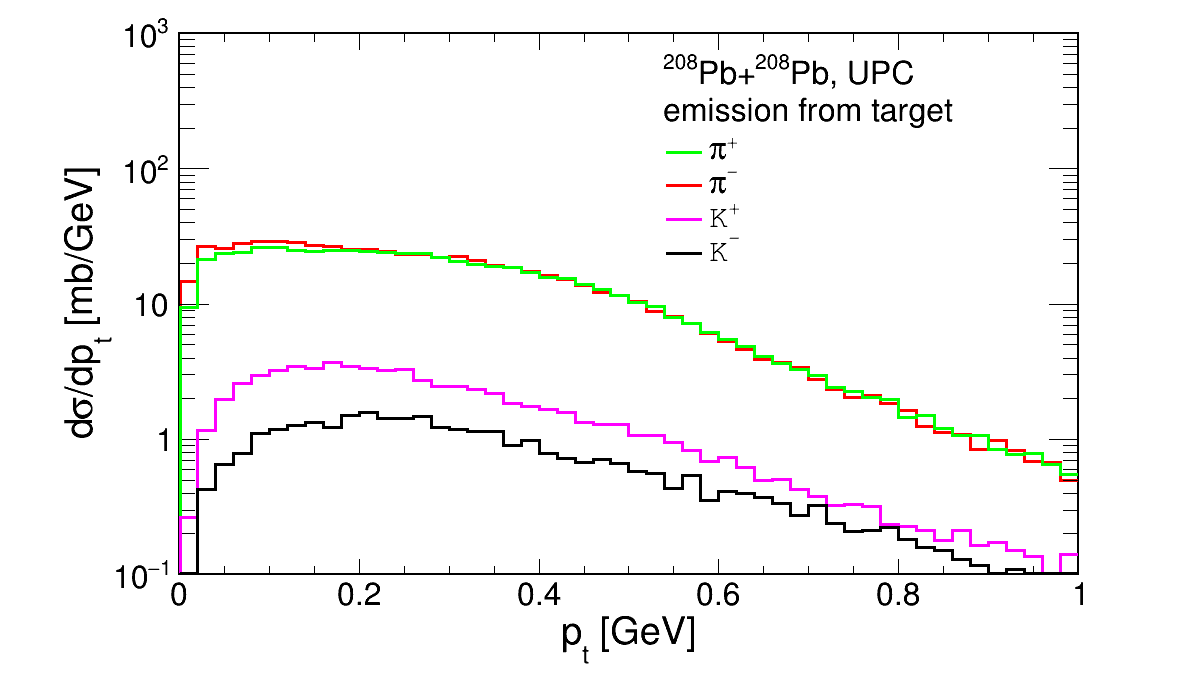}}
\put(4.7,8.7){\bf \large b)}

\put(-0.5,3){\includegraphics[width=0.54\textwidth]{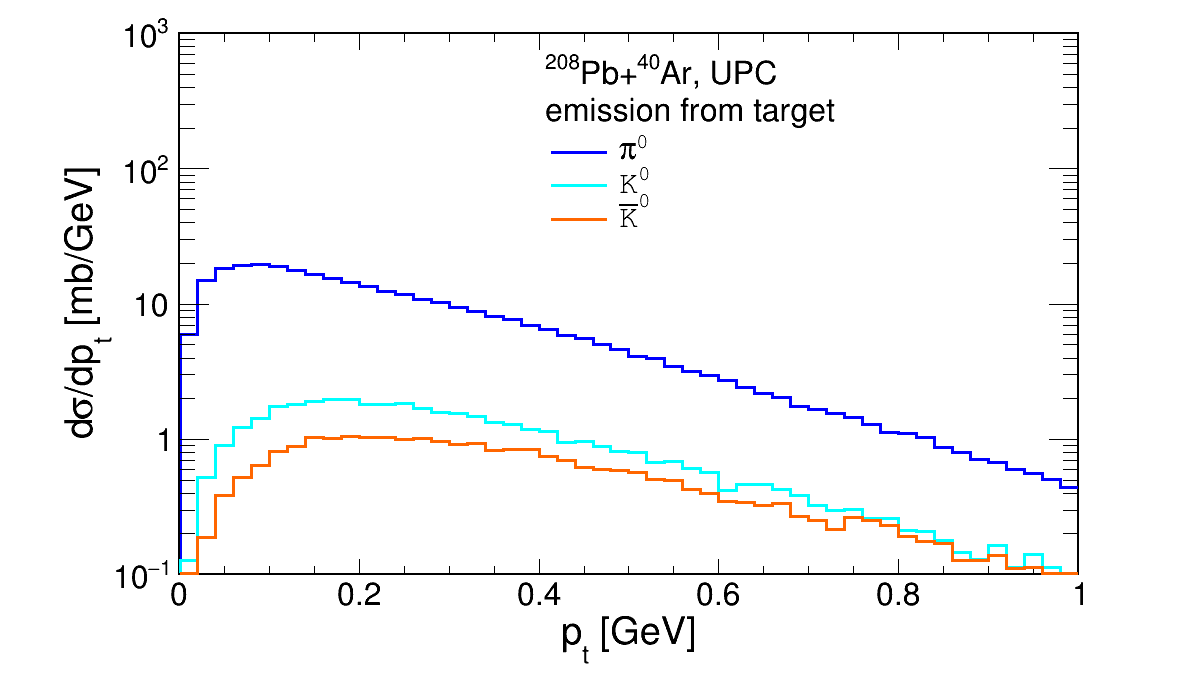}}
\put(-0.2,5.7){\bf \large c)}
\put(4.5,3){\includegraphics[width=0.54\textwidth]{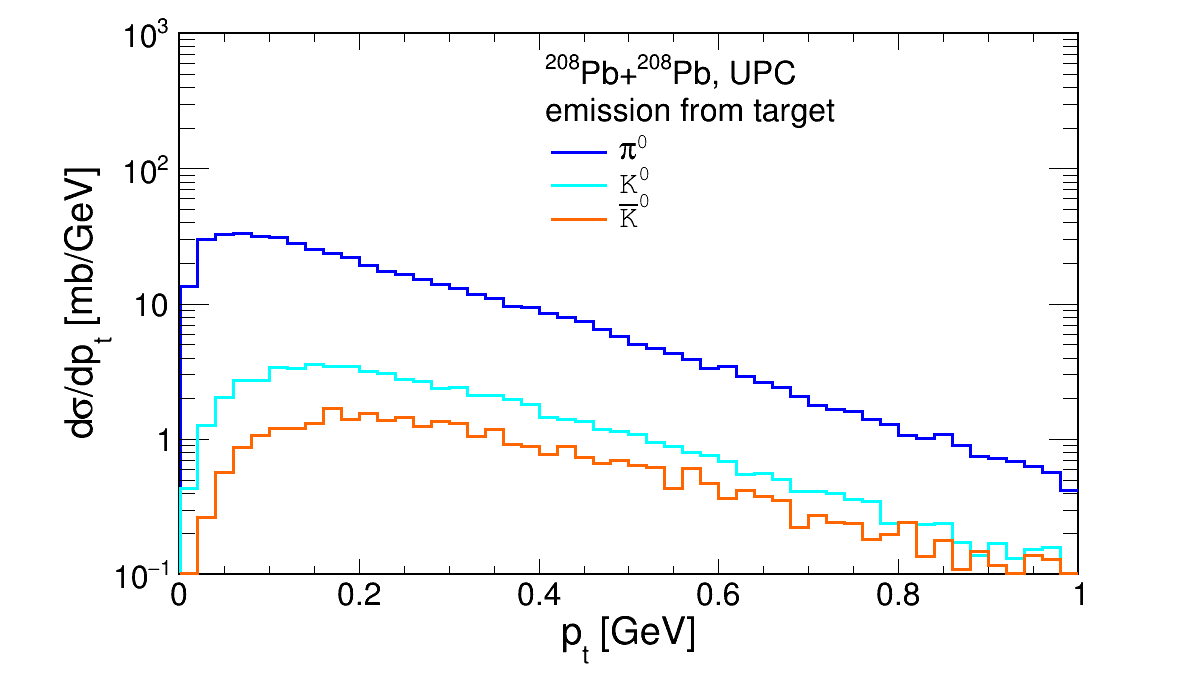}}
\put(4.7,5.7){\bf \large d)}
	\put(-0.5,0){\includegraphics[width=0.54\textwidth]{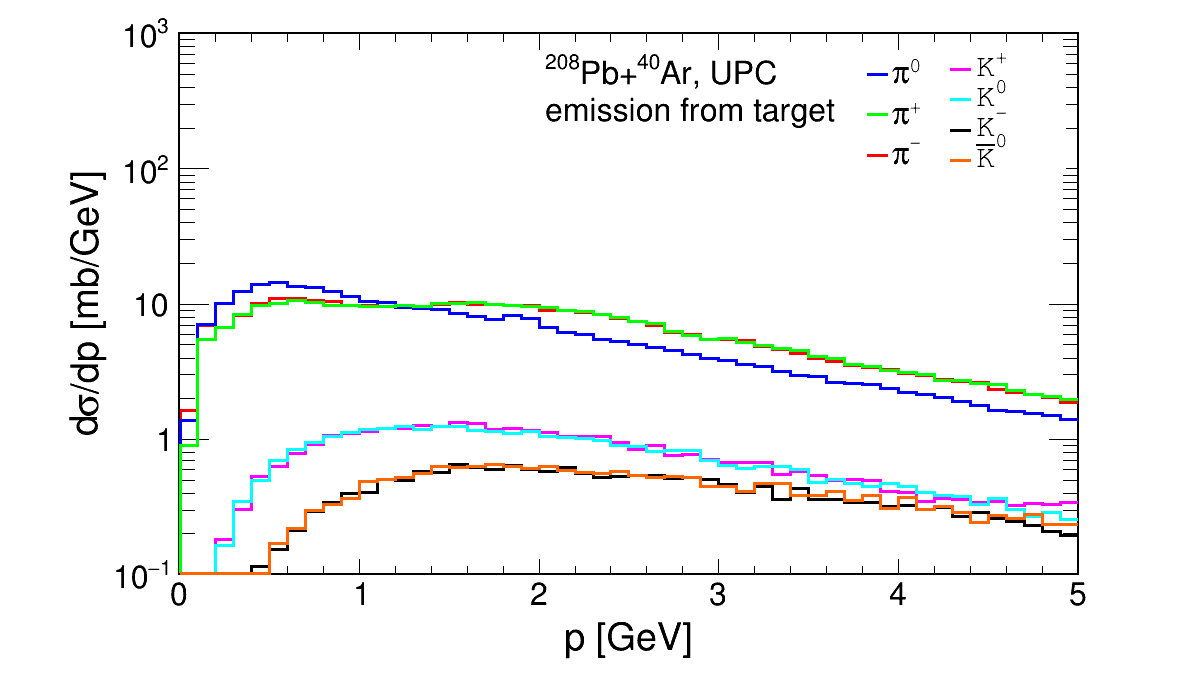}}
    \put(-0.2,2.7){\bf \large e)} 
	\put(4.5,0){\includegraphics[width=0.54\textwidth]{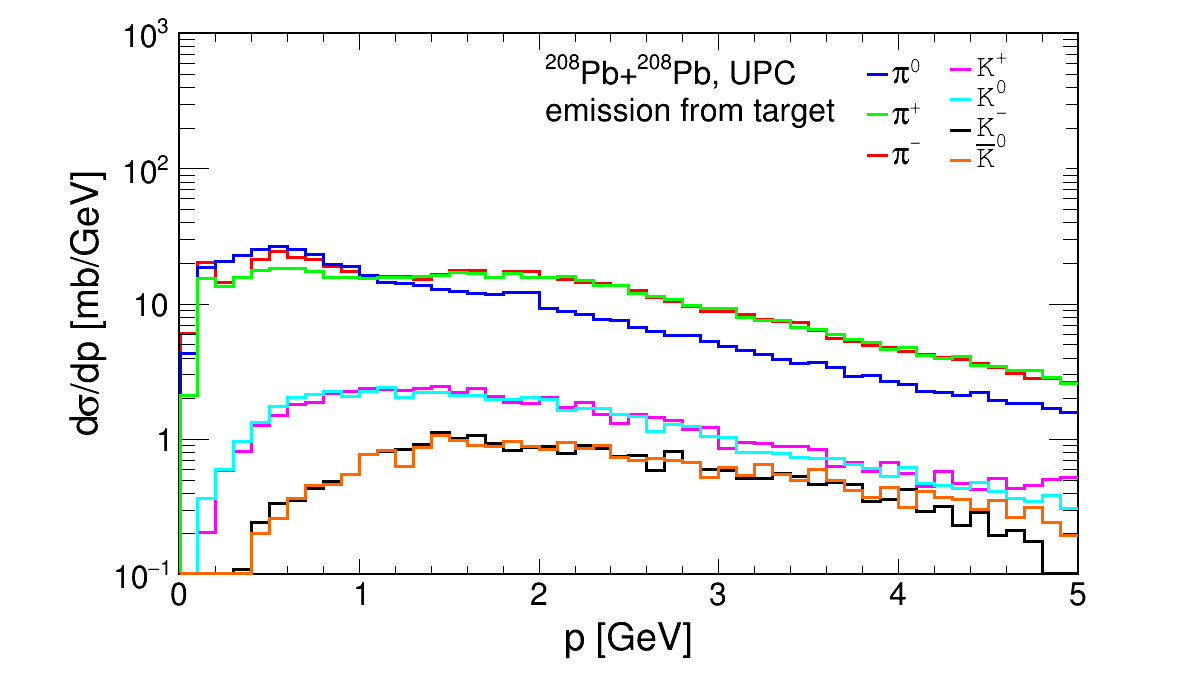}}
    \put(4.7,2.7){\bf \large f)} 
    \end{picture}
	\caption{
    Transverse momentum (upper and
medium panels) and momentum (lower panel) distributions of mesons produced in $^{208}\text{Pb}$+$^{40}\text{Ar}$ (left) and $^{208}\text{Pb}$+$^{208}\text{Pb}$ (right) 
 fixed-target collisions. The blue line represents $\pi^0$, the green line $\pi^+$, the red line $\pi^-$, the magenta line K$^+$, the cyan line K$^0$, the black line K$^-$, and the orange line $\bar{\text{K}}^0$. The results are obtained within the LHCb acceptance (2~$<$~$\eta$~$<$~5).
}
	\label{fig:LHCbcut_UPC_arpb}
\end{figure*}

Figure~\ref{fig:LHCb_UPC_ar_nuc} shows the transverse momentum and momentum distributions of neutrons and protons emitted by the excited $^{40}$Ar target nuclei restricted to the LHCb acceptance (2 $<$ $\eta$ $<$ 5). Compared with the corresponding full phase-space distributions shown in Fig.~\ref{fig:fullspace_UPC_ar_nuc}, the effect of the detector acceptance is more visible in the transverse momentum distribution than in the total momentum distribution.

The estimation of the nucleons emission with the LHCb pseudorapidity constraint is given in Tab.~\ref{tab:results_cut}. Even considering the restricted acceptance of the LHCb detector, the cross section for production of single proton or neutron is still rather large, of the order of 0.5~b for the $^{40}\text{Ar}$ target.

\begin{table}[!bt]
	\caption{Nucleons and mesons emission cross section from the $^{40}\text{Ar}$ (second column) and hypothetical $^{208}\text{Pb}$ (third column) target in UPC with the $^{208}\text{Pb}$ LHC beam. The maximal photon energy is $E_{\gamma, max}$ = 70 GeV. The results are calculated in the pseudorapidity range 2$<\eta<$5 with the GiBUU approach.}
	\label{tab:results_cut_bm}
	\centering
	\begin{tabular}{lccc}
		\hline\\
		 & $\quad$$^{208}$Pb+$^{40}$Ar [mb]$\quad$ &$^{208}$Pb+$^{208}$Pb [mb] $\quad$& $\sigma_{Pb}/\sigma_{Ar}$$\quad$\\\\
		\hline\\
		p			  & 461.3 $\pm$ 1.5 & 1248 $\pm$ 13 &  2.7\\
		n			  & 553.3 $\pm$ 1.8 & 3334 $\pm$ 29 & 6.0 \\\\
		$\pi^0$ 	  & 188.0 $\pm$ 1.2 & 435.0 $\pm$ 3.5 & 2.3 \\
		$\pi^+$ 	  & 280.6 $\pm$ 1.3 & 603.0 $\pm$ 3.9 & 2.2 \\
		$\pi^-$ 	  & 283.1 $\pm$ 1.3 & 636.4 $\pm$ 4.1 & 2.3 \\\\
		K$^+$ 	  		  & 34.52 $\pm$ 0.38 & 79.81 $\pm$ 0.96 & 2.4 \\
		K$^0$ 		  & 33.73 $\pm$ 0.39 & 78.13 $\pm$ 0.95 & 2.3 \\
		$\text{K}^-$ 	  & 19.88 $\pm$ 0.29 & 37.88 $\pm$ 0.59 & 1.9 \\
		$\bar{\text{K}}^0$   & 20.39 $\pm$ 0.30 & 38.43 $\pm$ 0.61 & 1.9 \\
        \\\hline\\
	\end{tabular}
\end{table}

Similarly, Fig.~\ref{fig:LHCbcut_UPC_arpb} shows the transverse momentum and momentum distributions of charged and neutral mesons emitted by the $^{40}$Ar target (left) and, for comparison, by the hypothetical $^{208}$Pb target (right), within the LHCb acceptance, i.e. requiring the pseudorapidity condition (2~$<$~$\eta$~$<$~5). Compared with the full phase-space distributions shown in Fig.~\ref{fig:fullspace_UPC_pb}, the $\eta$ requirement affects mainly the momentum distributions, in particular the low-$p$ region, especially for charged pions and kaons.

A detailed comparison of the cross section for emission of nucleons and mesons between $^{40}\text{Ar}$ target nuclei and hypothetical $^{208}\text{Pb}$ target nuclei within the LHCb acceptance is presented in Tab.~\ref{tab:results_cut_bm}. The probability of meson production is hindered in comparison to the emission of protons or neutrons. Moreover, the production of the charged pions is more probable than neutral ones. However, one should consider that, at present, no dedicated tools are available at LHCb for neutron detection and identification.
For the meson channels, the cross sections for the hypothetical $^{208}\text{Pb}$ target are approximately twice those for $^{40}$Ar, while for neutrons there are order of six  (see last column of Table~\ref{tab:results_cut_bm}).

Given the relatively high cross section values expected for the $^{40}$Ar target, even if restricted to the LHCb geometrical acceptance, these predictions look promising for a future measurement.

\section{Summary}
\label{Sect.VI}

In the present work, we have considered the production of charged and neutral particles (protons, neutrons, pions and kaons) from electromagnetic excitation of fixed-target $^{40}$Ar nuclei in UPC with a $^{208}$Pb LHC beam. Similar analyses with Ne, Kr and Xe targets are also possible at LHCb thanks to the SMOG2 fixed-target system.

The energy of the photon produced in ultraperipheral collisions depends on the nuclear charge, and its maximum value critically depends on the collision system and type. For instance this energy is of the order of $10^{5}$~GeV for $^{208}$Pb+$^{208}$Pb collisions in collider mode~\cite{Jucha2026}, while it is of order of 70~GeV for the case of fixed target $^{208}$Pb+$^{40}$Ar. 
Thus, the fixed-target mode provides the opportunity to
control the contribution of the high-energy component to the total cross section 
\cite{Jucha2025, Jucha2026}.

The LHCb apparatus, including the SMOG2 gaseous target system, opens the unique possibility to measure such processes in fixed-target UPC within the pseudorapidity range 2~$<\eta<$~5. 

The first exploratory calculation has been done using the GiBUU generator supplemented by the TCM proposed recently by our group. We have presented our predictions in the form of pseudorapidity, transverse momentum and momentum distributions. The results were complemented by tables with the corresponding fiducial cross sections. Rather large cross sections have been obtained for photoproduction of nucleons and mesons in UPC, ranging between milibarns and barns, i.e., in principle, easy to be measured by the LHCb experiment.

Despite the limited phase-space allowed by the LHCb spectrometer, the proposed measurement is of great importance and essentially unique. Thanks to the SMOG2 setup, the LHCb experiment is the only place worldwide where one can study ultraperipheral collisions at fixed-target kinematics with a heavy ion beam at the TeV scale. Furthermore, the availability of different target species, ranging from $^{4}$He to Xe, allows for the unique possibility to study the evolution of the photoproduction cross section for a variety of hadron types with the target mass.

\hspace{1cm}

\textbf{Acknowledgement}\\

\noindent Paweł Jucha acknowledges financial support provided by the Polish National Agency for Academic Exchange NAWA  under the Programme STER - Internationalisation of doctoral schools, Project no. BPI/STE/2023/1/00027/U/00001. We acknowledge support from University of Ferrara (Italy).



\end{document}